\documentclass[aps,pra,reprint,groupedaddress,showpacs]{revtex4-1}
\usepackage{amssymb}
\usepackage{amsmath}
\usepackage{dcolumn}
\usepackage{bm}
\usepackage{graphicx}
\usepackage{subfigure}
\usepackage{mathrsfs}
\usepackage{autobreak}
\usepackage[T1]{fontenc}
\usepackage[utf8]{inputenc}
\usepackage[colorlinks,linkcolor=blue,anchorcolor=blue,citecolor=blue,urlcolor=blue]{hyperref}
\allowdisplaybreaks
\begin{document}

\title{Magnon blockade in spin-magnon systems with frequency detuning }
\author{Sheng Zhao}
\author{Ya-Long Ren}
\author{Xin-Lei Hei}
\author{Xue-Feng Pan}
\author{Peng-Bo Li}
\email{lipengbo@mail.xjtu.edu.cn}
\affiliation{Ministry of Education Key Laboratory for Nonequilibrium Synthesis and Modulation of Condensed Matter, Shaanxi Province Key Laboratory of Quantum Information and Quantum Optoelectronic Devices, School of Physics, Xi'an Jiaotong University, Xi'an 710049, China}
\date{\today}
\begin{abstract}
Magnon blockade is a physical mechanism for the preparation of single-magnon source,
which has important applications in quantum information processing.
Here we propose a scheme for generating an optimal magnon blockade in the spin-magnon quantum system.
By introducing the frequency detuning between magnon and spin qubit of NV center, the conventional magnon blockade and the unconventional magnon blockade can
be obtained under both strong and weak coupling, relaxing the requirements for coupling
strength.
Moreover, the conventional and unconventional magnon blockade can occur simultaneously when both the magnon and the spin qubit are driven.
This allows the equal-time second-order correlation function to reach $10^{-8}$, about five orders of magnitude lower than that in previous works. Additionally, the time-delayed second-order correlation function avoids oscillation.
Our study demonstrates the impact of frequency detuning on the magnon blockade and proposes methods to enhance the magnon blockade and relax the requirements for coupling strength through frequency detuning.
\end{abstract}

\maketitle


\section{\label{sec:level1}INTRODUCTION  }

In recent years, hybrid quantum systems based on ferrimagnetic yttrium iron garnet (YIG) have received extensive attention in physics~\cite{knill2001scheme,kok2007linear,kimble2008quantum}. YIG~\cite{ gilleo1958magnetic,PhysRevLett.113.083603} has extremely high spin density, extremely low magnon damping rate, and nonlinear amplification and control of magnons~\cite{barker2016thermal, collet2016generation, princep2017full, wei2022giant, zhang2016cavity}, which allows the magnons of YIG to be strongly coupled with cavity photons~\cite{zhang2014strongly,huebl2013high, zare2015magnetic, maier2016spin, tabuchi2015coherent, li2016hyperparallel, clerk2020hybrid, soykal2010strong, bai2015spin, viola2016coupled, haigh2016triple, kounalakis2022analog}.
Based on the hybrid cavity‐magnonic system, many interesting phenomena have been studied, such as magnon-polariton bistability~\cite{wang2018bistability}, magnon dark-mode memory~\cite{zhang2015magnon, xiao2019magnon}, magnon non-reciprocity~\cite{iguchi2015nonreciprocal, kong2019magnon}, magnon Kerr effect~\cite{wang2018bistability, wang2016magnon}, magnon optical cooling~\cite{sharma2018optical}, and magnon quadrature squeezing~\cite{zhang2021generation,li2019squeezed}. In addition to the cavity magnonics, the magnons can also be directly coupled to the solid-state spin through the magnetic-dipole interaction.
As a typical solid-state spin, NV center with stable triple ground state is a good physical platform.
NV center exhibits strong decoupling from the environment~\cite{aharonovich2011diamond, barry2020sensitivity, doherty2013nitrogen, bar2013solid, doherty2014electronic, abobeih2018one}, so it has a low damping rate and a long coherence time,
which can be used for quantum storage~\cite{fuchs2011quantum}
and quantum sensing~\cite{kolkowitz2012coherent, dolde2011electric}.
Therefore, hybrid quantum system based on magnons and spin qubit of the NV center
may open new horizons for exploring quantum magnonics phenomena and designing future magnonics devices.

Recently, as a typical pure quantum effect, the investigation of the magnon blockade (MB) has drawn considerable attention in quantum information processing.
Analogous to the photon blockade~\cite{birnbaum2005photon,rabl2011photon,ridolfo2012photon} and phonon blockade~\cite{liu2010qubit,xie2017phonon,yao2022nonreciprocal},
MB is a phenomenon manifesting that the absorption of the first magnon will block subsequent magnons~\cite{liu2019magnon}. MB can be divided into the conventional magnon blockade (CMB) and the unconventional magnon blockade (UMB),
where the CMB is caused by the anharmonicity in the energy spectrum,
and the UMB is induced by the destructive quantum interference between different paths.
Many approaches have been proposed for generating the MB with different kinds of architectures,
such as a magnetic sphere indirectly coupled to a superconducting qubit mediated by a microwave cavity~\cite{ xie2020quantum},
a magnetic-cavity-mechanical system with an optical parametric amplifier~\cite{ zhang2024magnon},
a giant spin ensemble coupled to a waveguide~\cite{wang2024magnon},
and an optomagnonic system with chiral exceptional points~\cite{ yuan2023periodic}.
However, these studies mainly focused on resonant conditions where magnon and qubit frequencies match,
so the CMB is restricted to strong coupling~\cite{jin2023magnon,zhao2020simultaneous},
and the UMB requires weak coupling~\cite{wang2022hybrid, hou2024magnon},
which impose strict limitations on coupling strength.

In this paper, we theoretically investigate the MB in a hybrid system, where a magnon of YIG sphere is coupled directly to a spin qubit of NV center via magnetic-dipole interaction~\cite{hei2021enhancing}. To systematically investigate the mechanism for generating an optimal MB, various driving schemes have been employed. When only the magnon is driven, we can realize both the UMB and the CMB. The UMB can exist under both strong and weak coupling. Notably, UMB under strong coupling has a wider range than UMB under weak coupling. When the magnon and spin qubit are both driven, the CMB can exist even in the weak-coupling condition. Additionally, Adjusting the qubit-magnon frequency detuning enables the simultaneous occurrence of the CMB and the UMB. So, this allows the equal-time correlation function to reach $g^{\left( 2 \right)}\left( 0 \right) \sim 10^{-8}$ and the time-delayed correlation function $g^{\left( 2 \right)}\left( t \right)$ to avoid rapid oscillation. This approach is optimal as it combines the advantages of both the UMB and the CMB while avoiding their respective disadvantages. Our findings suggest an excellent way to achieve and enhance the MB by adjusting the qubit-magnon frequency detuning.

The rest of the work is organized as follows: In Sec.~\ref{2}, we derive the Hamiltonian of the spin-magnon hybrid system. In Sec.~\ref{3}, we study the physical mechanism for generating an optimal MB under various schemes. Finally, the conclusion is given in Sec.~\ref{4}. We provide details of the analytical solution to MB in the Appendix.

\section{Hamiltonian}\label{2}

\begin{figure}[tp]
	\centering
\includegraphics[width=1\columnwidth]{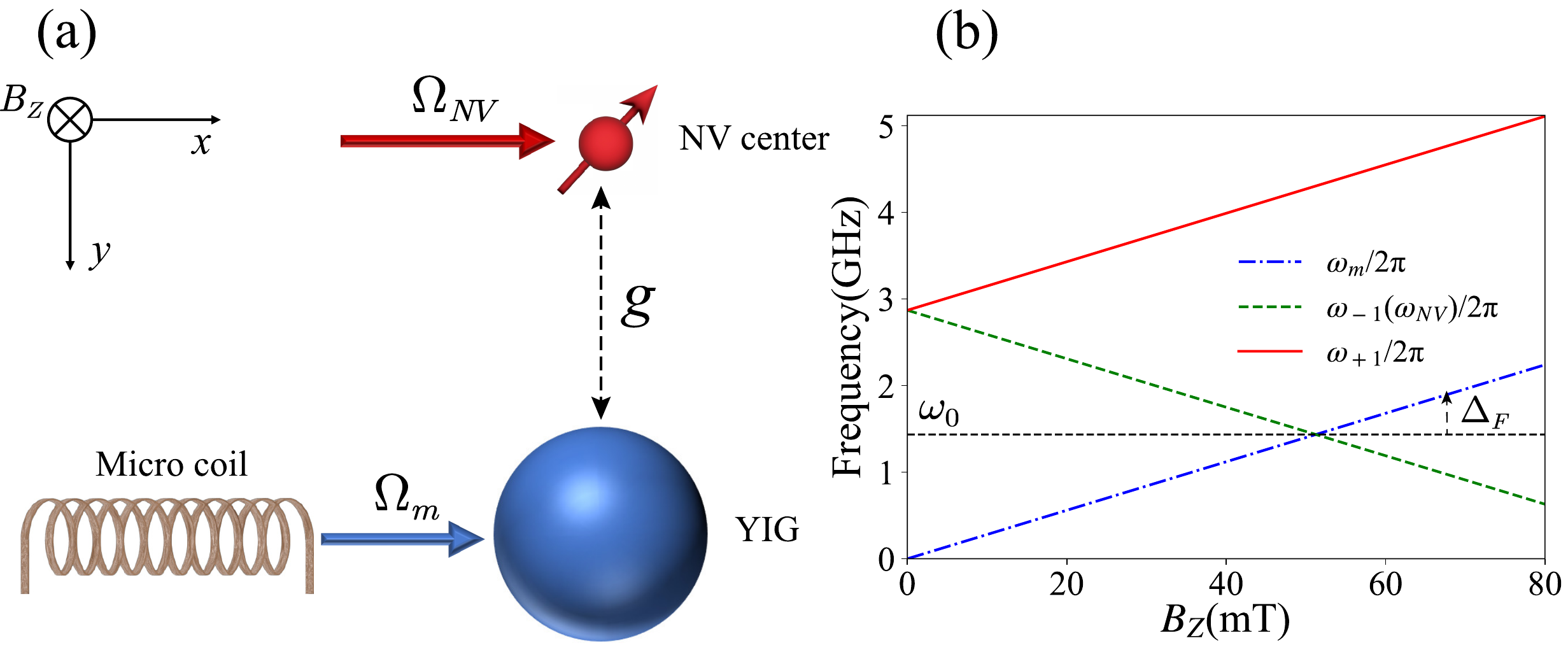}
\caption{\label{fig1}(a) Schematic diagram of the spin-magnon hybrid system, where a YIG sphere is directly coupled to the spin qubit of the NV center. A static magnetic field $B_Z$ oriented along the NV symmetry axis can be used to adjust the frequencies of the magnon and the NV center. The magnon and the spin qubit are under the driving fields with driving field strength $\Omega_m$ and $\Omega_\text{NV}$, respectively. (b) Plot of the frequencies of the magnon and the NV center as functions of the external magnetic field $B_Z$.}
\end{figure}

As shown in Fig.~\ref{fig1}(a), we consider a spin-magnon system in which a spherical micromagnet of radius $R$ is coupled to an NV center. In this system, the magnetic microsphere is a YIG sphere, and the NV center is placed on top of the YIG sphere. This YIG sphere provides spin waves, which exist in ferromagnetic or antiferromagnetic materials at a low energy limit. The spin wave in magnetic microspheres can be quantized as the magnon by dipolar, isotropic, and magnetostatic approximations~\cite{gonzalez2020theory}. In our system, only Kittel mode in the YIG sphere is considered, where all spins are in phase and have the same amplitude~\cite{kittel1948theory, tabuchi2016quantum}. The free Hamiltonian of the magnon with the annihilation (creation) operator $m$ $(m^{\dag})$ is expressed as (we set $\hbar= 1$)
\begin{equation}
H_m=\omega _mm^{\dag}m,
\end{equation}
where the resonance frequency of the magnon is $\omega _m=\left|\gamma_{e}\right|B_Z,$ with the gyromagnetic ratio $\gamma_{e}/2\pi=28$ GHz/T,
which can be controlled by the external magnetic field $B_Z$.

For the NV center, it is the triple spin of total spin $S=1$. And the eigenstates of spin operator $S_Z$ are $\left\{  \left| 0 \right> , \left| \pm 1 \right>\right\}$ with $S_Z\left| j \right>=j\left| j \right>$ and $j = 0$ and $\pm1$. Since the external magnetic field $B_Z$ exists along the axis of symmetry of the NV center, the degeneracy of state $\left| \pm 1 \right>$ splits by the Zeeman effect~\cite{doherty2013nitrogen,li2019interfacing} with $\omega _{\pm}=D_0\pm \left| \gamma_{e} \right|B_Z$ taking $\left| 0 \right>$ as the energy reference value. Here, $D_0=2\pi \times 2.87$~\text{GHz} is the zero-field splitting. We plot the frequency split of the NV center and the magnon resonance frequency changed with the external magnetic field in Fig.~\ref{fig1}(b). It can be found that the state  $\left| +1 \right>$ can be safely excluded due to its off-resonance with the magnon~\cite{hei2021enhancing}. Therefore, we can focus on the states $\left| 0 \right>$ and $\left| -1 \right>$, which together encode a two-level spin qubit. The free Hamiltonian of the spin qubit of the NV center can be expressed as
\begin{equation}
H_\text{NV}= \omega _\text{NV}\sigma _+\sigma_-,
\end{equation}
where $\sigma _-$ $(\sigma _+)$ is the spin lowering (raising) operator.
Here, we have taken $\omega _\text{NV}=\omega _{-1}$.

The magnon of the YIG sphere could be directly coupled to the spin qubit of the NV center through the magnetic-dipole interaction~\cite{hei2021enhancing}. Assuming that the coupling strength is less than the resonance frequency,
by using the rotation-wave approximation, the spin-magnon interaction can be described by the Hamiltonian
\begin{equation}
H_\text{int}= g\left( m\sigma _++m^{\dag}\sigma _- \right),
\end{equation}
where $g$ is the spin-magnon coupling strength.
When the external magnetic field $B_Z$ is tuned to $\frac{D_0}{2\left| \gamma_{e} \right|}$, the frequencies of the magnon and the spin qubit of the NV center can be matched. Without loss of generality, we can express the frequencies of the magnon and the spin qubit of the NV center as $\omega _m=\omega _0+\Delta _F$ and $\omega _\text{NV}=\omega _0-\Delta _F$, respectively. Here, $\omega _0=D_0/2$, and $\Delta _F=\left| \gamma_{e} \right|\left( B_Z-\frac{D_0}{2\left| \gamma_{e} \right|} \right)$ represents half of the frequency detuning between the magnon and the spin qubit of the NV center.

To realize and manipulate the MB, we can apply two microwave fields to drive the magnon and the spin qubit.
Then, the total Hamiltonian of the spin-magnon system is given by
\begin{align}
H_\text{tot}=& \left( \omega _0+\Delta _F \right) m^{\dag}m+  \left( \omega _0-\Delta _F \right) \sigma _+\sigma _-\nonumber\\
&
+  g\left( m\sigma _+
+m^{\dag}\sigma _- \right)
+ \Omega _m\left( m^{\dag}e^{-i\omega_l t}+me^{i\omega_l t} \right)\nonumber\\
&+ \Omega _\text{NV}\left( \sigma _+e^{-i\omega_l t}
+\sigma _-e^{i\omega_l t} \right),
\end{align}
where $\Omega_m$ ($\Omega _\text{NV}$) is the Rabi frequency for driving the magnon (spin qubit). The two microwave fields are assumed to have the same frequency $\omega_l$. In the rotating frame with respect to $V=\omega _l\left( m^{\dag}m+\sigma _+\sigma _- \right)$, we can remove the time dependence of the total Hamiltonian. Consequently, the effective Hamiltonian of the spin-magnon system can be obtained as
 \begin{align}
H_\text{eff}&=( \Delta +\Delta _F ) m^{\dag}m+( \Delta -\Delta _F ) \sigma _+\sigma _-\nonumber\\
&+g( m\sigma _++m^{\dag}\sigma _- ) +\Omega _m( m^{\dag}+m ) +\Omega _\text{NV}( \sigma _++\sigma _-)\label{eq:5}, \nonumber\\
 \end{align}
where $\Delta=\omega_0-\omega_l$ is the driving detuning.

\section{Magnon blockade}\label{3}
\begin{figure*}[htbp]
\includegraphics[width=1.8\columnwidth]
{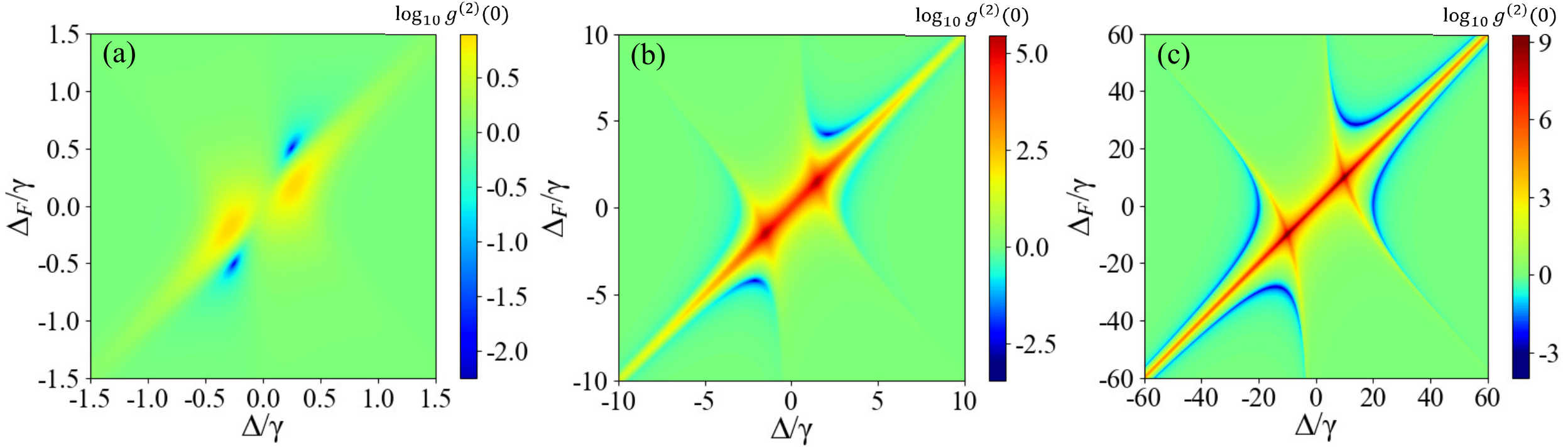}
\caption{\label{fig2} The equal-time second-order magnon correlation function $\log_{10} g^{\left( 2 \right)}\left( 0 \right)$ as a function of driving detuning $\Delta$ and frequency detuning $\Delta_F$ under the magnon drive. The coupling strength is chosen as (a) $g/\gamma=0.5$, (b) $g/\gamma=3$, (c) $g/\gamma=20$. Here, $\gamma=2\pi\times1$ MHz is used as the frequency unit. The selection of the remaining parameters is as follows: $\Omega_m/\gamma=0.01$, $\kappa/\gamma=0.5$, $n_{th}=0$.}
\end{figure*}

Since the system is an open quantum system, the operators in the microscopic system will be coupled with the variables of the macroscopic system, resulting in the system's dissipation. Therefore, we use the quantum master equation to study the dynamical evolution of the system. The mast equation of the system can be expressed as
 \begin{align}
\frac{\partial}{\partial t}\rho =&-i\left[ H_\text{eff},\rho \right] +\frac{\kappa _m}{2}\left( n_{th}+1 \right) \mathcal{L}_m\left[ \rho \right]\nonumber\\
&+\frac{\kappa _m}{2}n_{th}\mathcal{L}_{m^{\dag}}\left[ \rho \right] +\frac{\kappa _q}{2}\mathcal{L}_{\sigma _-}\left[ \rho \right]\label{eq:6},
\end{align}
where $\rho$ is the density matrix of the system, $n_{th}$ is the equilibrium thermal magnon occupation number, and $\mathcal{L}_o\left[ \rho \right] =2o\rho o^{\dag}-o^{\dag}o\rho -\rho o^{\dag}o$ is the Lindblad superoperators for a given operator $o$ ($o=m,\sigma_-$). $\kappa _m$ and $\kappa _q$ are the decay rates of the magnon and the spin qubit, respectively.
In the following, we take $\kappa _m=\kappa _q=\kappa$ for simplicity. To investigate the MB, it is necessary to study the quantum statistical properties of magnon, which can be well characterized by using the equal-time second-order correlation function
\begin{equation}
g^{\left( 2 \right)}\left( 0 \right) =\frac{\text{Tr}\left( \rho m^{\dag}m^{\dag}mm \right)}{\left[ \text{Tr}\left( \rho m^{\dag}m \right) \right] ^2}=\frac{\left< m^{\dag}m^{\dag}mm \right>}{\left< m^{\dag}m \right> ^2}. \label{eq:7}
\end{equation}
In this paper, we will use the QUTIP package in PYTHON to numerically simulate the statistical properties of the magnon~\cite{johansson2012qutip}.
$g^{\left( 2 \right)}\left( 0 \right)>1$ reveals that when one magnon is excited, the probability of another magnon being excited in a short time interval is high, so the magnon tends to cluster together, which is called the bunching effect of magnon. It corresponds that the number of magnons satisfies the super-Poisson distribution. $0<g^{\left( 2 \right)}\left( 0 \right)<1$ reveals that when one magnon is excited, it is difficult to excite another magnon in a short interval of time, so the magnon tends to appear independently, which is called the anti-bunching effect of magnon. It corresponds that the number of magnons satisfies the sub-Poisson distribution. Especially, the limit $g^{\left( 2 \right)}\left( 0 \right) \rightarrow 0$ indicates the MB~\cite{liu2019magnon,jin2023magnon}, and $g^{\left( 2 \right)}\left( 0 \right)\gg1$ indicates the magnon-induced tunneling (MIT)~\cite{xu2013photon,kowalewska2019two,zhai2019mechanical}. Subsequently, we will show how to realize the MB under different driving conditions.

\subsection{Driving the magnon}
Now, we consider that only the magnon is driven with $\Omega _\text{m}\neq0$ and $\Omega _\text{NV}=0$. As shown in Fig.~\ref{fig2},
we plot the equal-time second-order correlation function $\log_{10}g^{\left( 2 \right)}\left( 0 \right)$ as a function of the
the driving detuning $\Delta$ and the frequency detuning $\Delta _F$ by numerically solving the Eq.~(\ref{eq:7}).
Here, we consider weak coupling $g/\gamma=0.5$ in (a), medium coupling $g/\gamma=3$ in (b), and strong coupling $g/\gamma=20$ in (c).
Under the weak coupling, MB appears as two centrally symmetrical points.
As the coupling strength further increases, MB appears as a pair of hyperbola and a pair of center-symmetric curves.

To better explain these numerical results, we now present the analytical solutions. At the semiclassical limit, the effect of quantum jumps can be neglected, so the quantum master equation can be well described via Schrödinger equation $i\partial \left| \psi \right> /\partial t=H_\text{non}\left| \psi \right>$ of non-Hermitian Hamiltonian~\cite{plenio1998quantum,minganti2019quantum}
\begin{equation}
H_\text{non}=H_\text{eff}-\frac{i\kappa}{2}\left( m^{\dag}m+\sigma _+\sigma _- \right).
\end{equation}
In the weak driving field limit ($\Omega _\text{m}, \Omega _\text{NV}\ll\kappa$), the Hilbert space of the system can be approximately truncated within the two excitation subspaces. So the wave function $\left| \psi \right>$ of the system can be described as $\left| \psi \right> =C_{0g}\left| 0,g \right> +C_{0e}\left| 0,e \right> +C_{1g}\left| 1,g \right> +C_{1e}\left| 1,e \right> +C_{2g}\left| 2,g \right> $. Here $|C_{ng(e)}|^2$ represents the probability of detecting the direct product state $\left| n,g(e) \right>=\left| n \right>\otimes\left| g(e) \right>$, with $n$ magnons and spin qubit in the ground (excited) state. Substituting the wave function $\left| \psi \right>$ into the Eq.~(\ref{eq:7}), the second-order correlation function $g^{\left( 2 \right)}\left( 0 \right)$ can be expressed as [see Appendix~\ref{A1} for more detail]
\begin{equation}
g^{\left( 2 \right)}\left( 0 \right)\approx \frac{2\left| C_{2g} \right|^2}{\left| C_{1g} \right|^4}
=\frac{\left| B_1 \right|^2| C |^2 }{\left| A_1 \right|^2| D|^2}.\label{eq:9}
\end{equation}
where
\begin{align}
|C_{1g}|^2=&\frac{\Omega _{m}^{2}\left| A_1 \right|}{\left| C \right|^2},\quad
|C_{2g}|^2=\frac{\Omega _{m}^{4}\left| B_1 \right|^2}{2\left| C \right|^2\left| D \right|^2},\nonumber\\
\left| A_1 \right|^2=&\left[ \left( \Delta -\Delta _F \right) ^2+\kappa ^2/4 \right] ^2,\nonumber\\
\left| B_1 \right|^2=&\left( -2\Delta ^2+2\Delta _F\Delta -g^2+\kappa ^2/2 \right) ^2+\kappa ^2\left( 2\Delta -\Delta _F \right) ^2.\nonumber\\
\left| C \right|^2=&\left( g^2+\Delta _{F}^{2}-\Delta ^2+\kappa ^2/4 \right) ^2+\kappa ^2\Delta ^2,\nonumber\\
\left| D \right|^2=&\left( 2\Delta ^2+2\Delta _F\Delta -g^2-\kappa ^2/2 \right) ^2+\kappa ^2\left( 2\Delta +\Delta _F \right) ^2. \nonumber\\
\label{eq:11}
\end{align}
From Eq.~(\ref{eq:9}), we can obtain $g^{\left( 2 \right)}\left( 0 \right)\propto{\left| B_1 \right|^2\left| C \right|^2}$.
So, to generate the MB, we should keep $|B_1|^2\rightarrow0$, $|C|^2\rightarrow0$,
where $|B_1|^2\rightarrow0$ refers to the UMB, and $|C|^2\rightarrow0$ corresponds to the CMB.
First, we focus on the strong-coupling condition with $g\gg\kappa$.
\begin{figure}[tp]
	\centering
\includegraphics[width=1\columnwidth]{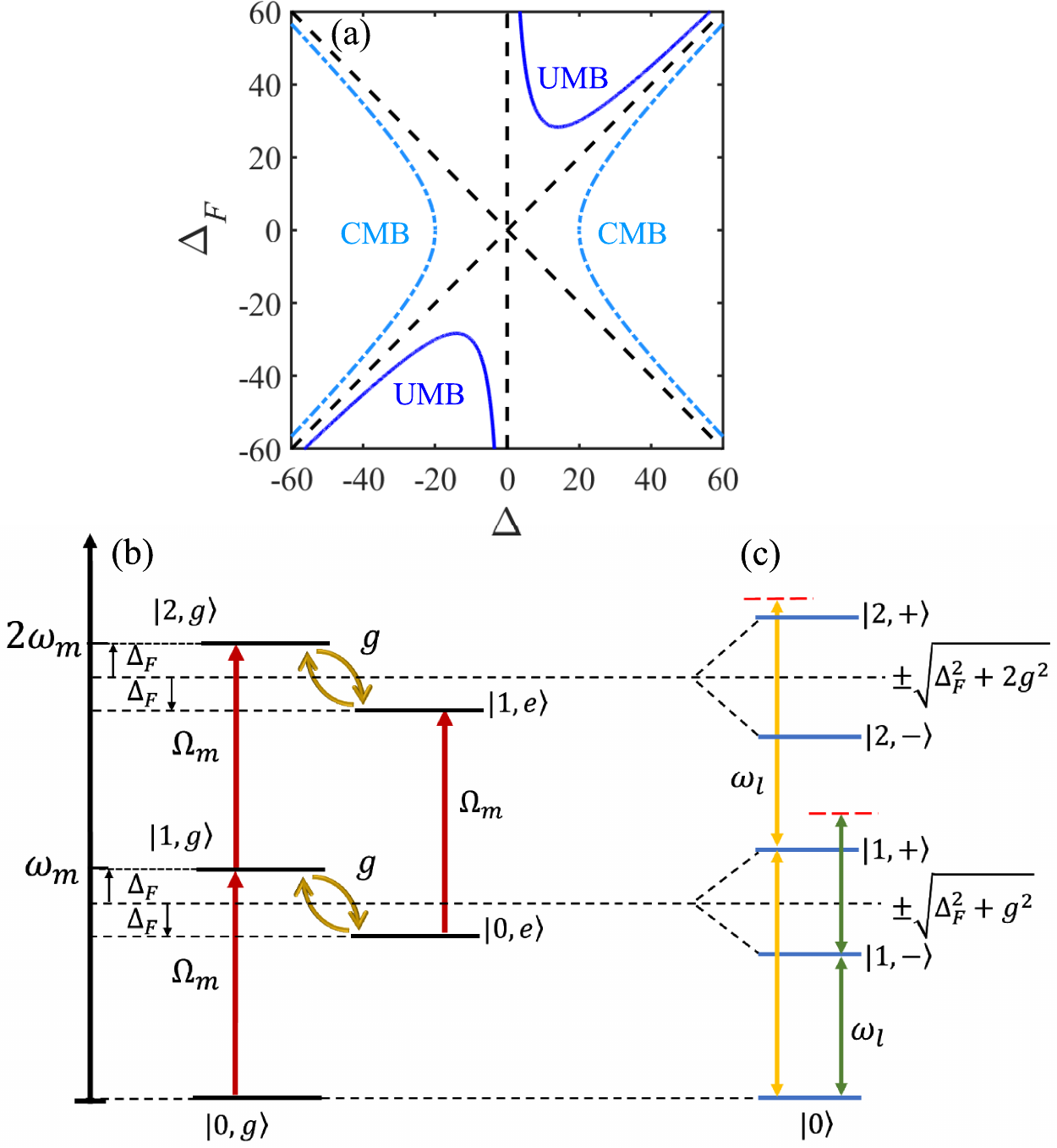}
\caption{(a) Structure of analytical solution in terms of the CMB (light blue), the UMB (dark blue) and the asymptote (dashed) when the magnon is driven. Energy-level diagram for generating (b) the UMB and (c) the CMB.
	\label{fig3}}
\end{figure}
When $| B_1 |^2\rightarrow0$, which is equivalent to
\begin{equation}
\Delta =\Delta _F/2\pm \sqrt{\Delta _{F}^{2} /4-g^2/2},
\label{eq:11}
\end{equation}
this solution is marked by two dark blue lines in Fig.~\ref{fig3}(a).
Physically, this can be understood by the destructive quantum interference between different quantum transition pathways,
as shown in Fig.~\ref{fig3}(b).
Obviously, there are two transition pathways corresponding to the transition from $\left| 0,g \right>$ to $\left| 2,g \right>$.
Here, first one is the direct excitation from $\left| 0,g \right>$ to $\left| 2,g \right>$,
and the other one is the coupling-mediated transitions $\left| 0,g \right>\rightarrow\left| 1,g \right>
\rightarrow\left| 0,e \right>\rightarrow\left| 1,e \right>\rightarrow\left| 2,g \right>$.
Owing to the destructive quantum interference of the two transition pathways,
\begin{figure}[tp]
	\centering
\includegraphics[width=0.8\columnwidth]{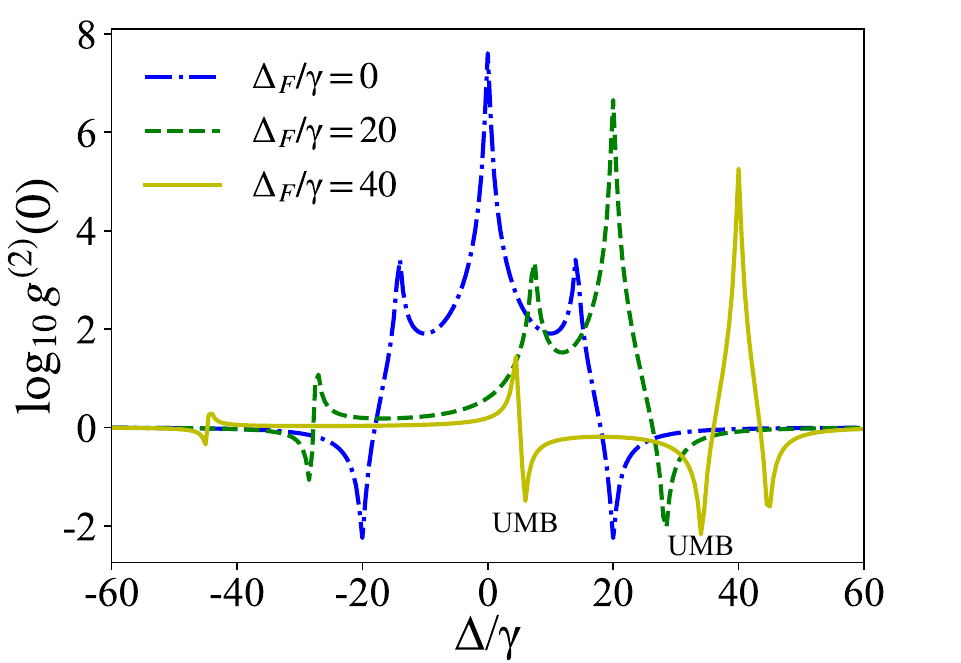}
\caption{Graph of the relationship between the second-order correlation function $\log_{10} g^{\left( 2 \right)}\left( 0 \right)$ and the driving detuning $\Delta$ under different frequency detuning $\Delta_F$, where we take $g/\gamma=20$ and $\Omega_m/\gamma=0.01$. The other parameters are the same as those in Fig.~\ref{fig2}. \label{fig4}}
\end{figure}
the two-magnon excitation is suppressed with $|C_{2g}|^2\rightarrow0$, leading to the UMB.
This suggests that the frequency detuning $\Delta_F$ is critical to the emergence of the UMB.

When $|C|^2\rightarrow0$, we have
\begin{equation}
 \Delta =\pm \sqrt{g^2+\Delta _{F}^{2}},
 \label{eq:12}
\end{equation}
which is a standard hyperbola.
Figure.~\ref{fig3}(a) display the standard hyperbola $(\Delta/g)^2-(\Delta_F/g)^2=1$
marked by two light blue lines.
On these two hyperbola lines, we can keep $|C|^2\rightarrow0$ and thus $g^{\left( 2 \right)}\left( 0 \right) \rightarrow 0$. 
Physically, the CMB can be attributed to the strong anharmonicity of the dressed states as seen in Fig.~\ref{fig3}(c). 
When the driving field satisfies $\omega_m-\Delta_F-\omega_l=\pm \sqrt{g^2+\Delta _{F}^{2}}$, 
the transitions from the ground state $\left| 0,g \right>$ to the dressed states
$\left| 1,\pm \right>$ are allowed. 
However, the transitions from $\left| 1,\pm \right>$ to $\left| 2,\pm \right>$ are suppressed 
due to $2\omega_m-\Delta_F-2\omega_l\ne\pm\sqrt{\Delta_F^2+2g^2}$.
This represents the CMB caused by non-equidistant energy levels.

Second, let us study the effects of the decay rates of the magnon and the spin qubit on the MB.
In this case, the conditions of $g,\Delta,\Delta_F\gg\kappa$ cannot be satisfied, and thus the CMB fails to appear.
This is in sharp contrast to the UMB.
When  $\Delta=\pm\sqrt{2g^2-\kappa^2}/2$ and $\Delta_F=2\Delta$ are both satisfied,
we can still keep $| B_1 |^2=0$.
So, we can achieve the UMB in the case of weak coupling,
which is consistent with the results given by the numerical simulations (see Fig.~\ref{fig2}(a)).

\begin{figure*}[htbp]
\includegraphics[width=1.8\columnwidth]
{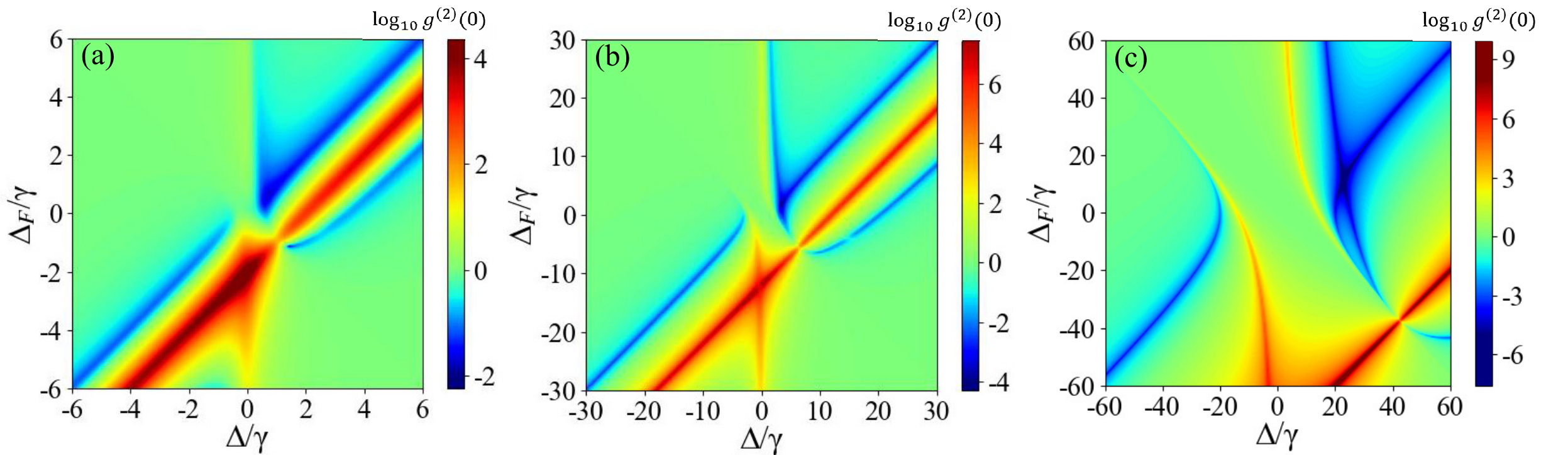}
\caption{\label{fig5} The equal-time correlation function $\log_{10} g^{\left( 2 \right)}\left( 0 \right)$ versus the driving detuning $\Delta$ and the frequency detuning $\Delta_F$ with $\lambda=4$ for (a) $g/\gamma=0.5$, (b) $g/\gamma=3$, (c) $g/\gamma=20$ under both the magnon and the qubit drives. The other parameters are the same as those in Fig.~\ref{fig2}. }
\end{figure*}

Finally, in order to demonstrate more clearly the effects of the driving detuning $\Delta$ and the frequency detuning $\Delta_F$ on the equal-time second-order correlation function $g^{\left( 2 \right)}\left( 0 \right)$, we display $\log_{10} g^{\left( 2 \right)}\left( 0 \right)$ as a function of the driving detuning $\Delta$ under different $\Delta_F$ in Fig.~\ref{fig4}. For $\Delta_F=0~\text{or}~20$, only the CMB exists, while the UMB appears for $\Delta_F=40$, which is consistent with the solution of the UMB in Eq.~(\ref{eq:11}). More importantly, the UMB under strong coupling has a wider range, which is different from the UMB under weak coupling disappearing with small changes in parameters \cite{wang2022hybrid, hou2024magnon}.

\subsection{Driving the magnon and the spin qubit}

\begin{figure}[tp]
	\centering
    \includegraphics[width=1\columnwidth]{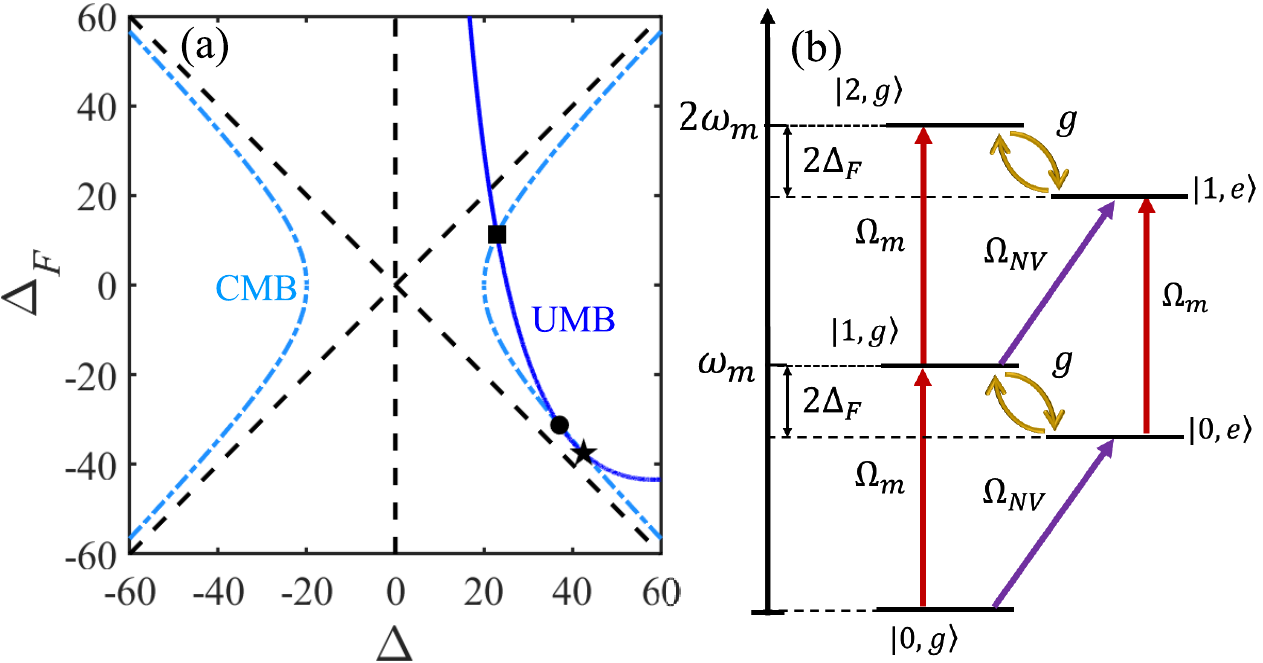}
\caption{(a) Structure of analytical solution in terms of the CMB (light blue), the UMB (dark blue), and the asymptote (dashed) when both the magnon and the qubit are driven. The pentagram, circle, and rectangle represent intersection point $\Delta_F(1)$, $\Delta_F(2)$, and $\Delta_F(3)$ respectively.
(b) The transition pathways when both the magnon and the qubit are driven.
	\label{fig6}}
\end{figure}

In this subsection, we show how to implement both the CMB and the UMB to achieve an optimal MB.
To this end, we consider the case where two microwave fields are applied to drive the magnon and spin qubit, respectively.
As shown in Fig.~\ref{fig5}, we plot the equal-time second-order correlation function $\log_{10} g^{\left( 2 \right)}\left( 0 \right)$ as a function of the driving detuning $\Delta$ and the frequency detuning $\Delta_F$ under weak coupling $g/\gamma=0.5$, medium coupling $g/\gamma=3$ and strong coupling $g/\gamma=20$ by numerically solving the Eq.~(\ref{eq:7}). We can observe that the CMB exists even in the case of weak coupling and gradually becomes stronger as the coupling strength increases. This is because the MIT is downward-offset to stay away from the CMB compared to the single-magnon drive scheme. Meanwhile, the UMB appears to be shifted downward compared to the single magnon-driven scheme.
To better understand these numerical results, we derive the equal-time second-order correlation function $g^{\left( 2 \right)}\left( 0 \right)$
in the case of two driving fields [see Appendix~\ref{B1} for more detail]
\begin{equation}
g^{\left( 2 \right)}\left( 0 \right)\approx \frac{2\left| C_{2g}^{\prime} \right|^2}{\left| C_{1g}^{\prime} \right|^4}
=\frac{\left| B_2 \right|^2| C |^2 }{\left| A_2 \right|^2| D|^2}.\label{eq:13}
\end{equation}
where
\begin{align}
|C_{1g}^{\prime}|^2&=\frac{\Omega _{m}^{2}\left| A_2 \right|}{\left| C \right|^2},\quad
|C_{2g}^{\prime}|^2=\frac{\Omega _{m}^{4}\left| B_2 \right|^2}{2\left| C \right|^2\left| D \right|^2},\nonumber\\
\left| A_2 \right|^2&=\left[ \left( \Delta -\Delta _F-\lambda g \right) ^2+\kappa ^2/4 \right] ^2,\nonumber\\
\left| B_2 \right|^2&=\left[ -2\left( \Delta ^2-\kappa ^2/4 \right) +\left( 4\lambda g+2\Delta _F \right) \Delta -\left( 1+\lambda ^2 \right) g^2 \right] ^2\nonumber\\
&+\kappa ^2\left[ 2\Delta -2\lambda g-\Delta _F \right] ^2.
\end{align}
Here, the parameter $\lambda=\Omega_\text{NV}/{\Omega_m}$ is the ratio of the Rabi frequency of the spin qubit to that of the magnon.
From Eq.~(\ref{eq:13}), we can obtain $g^{\left( 2 \right)}\left( 0 \right)\propto{\left| B_2 \right|^2\left| C \right|^2}$,
meaning that the CMB is independent of the driving fields, while the UMB is greatly affected by the driving fields.
When $\kappa\rightarrow0$, $| B_2 |^2\rightarrow0$ is equivalent to
\begin{equation}
\Delta=\Delta _F/2+\lambda g\pm \sqrt{\left( \Delta _F/2+\lambda g \right) ^2-\left( 1+\lambda ^2 \right)g^2/2}.
\label{eq:15}
\end{equation}
In Fig.~\ref{fig6}(a), we elucidate the relationship between $\Delta$ and $\Delta_{F}$,
which is marked by two dark blue lines.
When $\Delta_F$ approaches infinity, this pair of curves becomes
$\Delta\approx \Delta _F+2\lambda g$ or $\Delta\approx 0$.
Physically, this is due to the influence of the qubit-driving-induced transition pathways
$\left| 0,g \right>\rightarrow\left| 0,e \right>$ and $\left| 1,g \right>\rightarrow\left| 1,e \right>$ in Fig.~\ref{fig6}(b),
so that the UMB appears as a downward shift of $2\lambda g$.
This is completely different from the single magnon-driven scheme, in which the CMB and the UMB do not occur simultaneously.

\begin{figure*}[htbp]
\includegraphics[width=1.8\columnwidth]{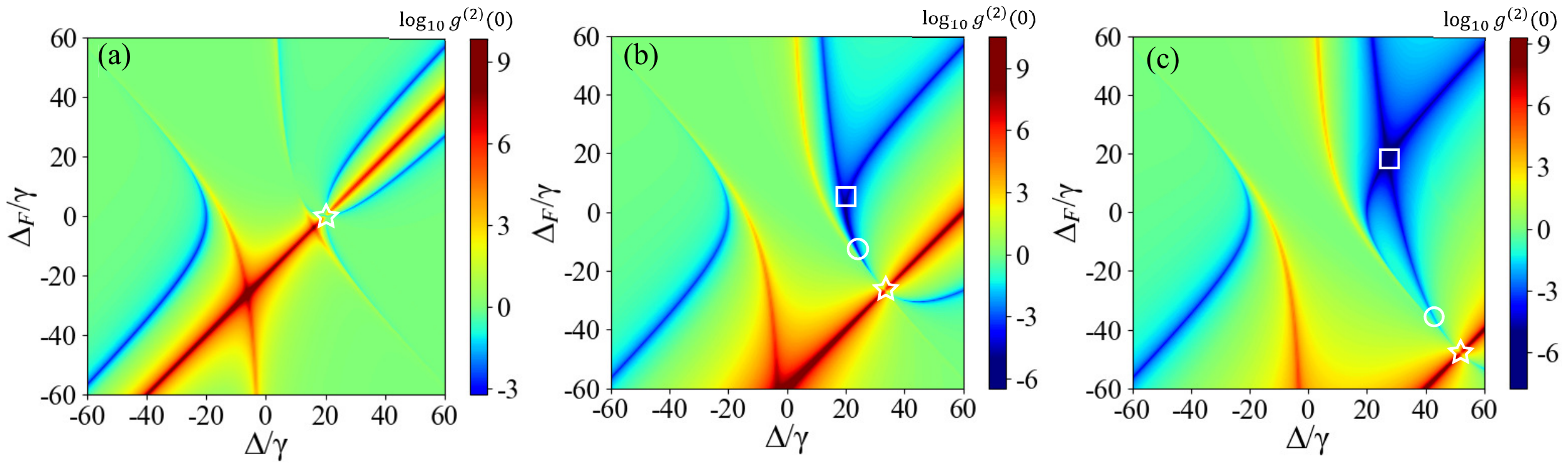}
\caption{correlation function $\log_{10} g^{\left( 2 \right)}\left( 0 \right)$ versus the driving detuning $\Delta$ and the frequency detuning $\Delta_F$ for $g/\gamma=20$ under both the magnon and the qubit drives. Here, the ratio of the Rabi frequency of the spin qubit to that of the magnon is chosen as (a) $\lambda=1$, (b) $\lambda=3$, and (c) $\lambda=5$. The pentagram, circle, and rectangle represent intersection point $\Delta_F(1)$, $\Delta_F(2)$, and $\Delta_F(3)$ respectively. The other parameters are the same as those in Fig.~\ref{fig2}.
	\label{fig7}}
\end{figure*}

\begin{figure}[htbp]
	\centering
\includegraphics[width=0.8\columnwidth]{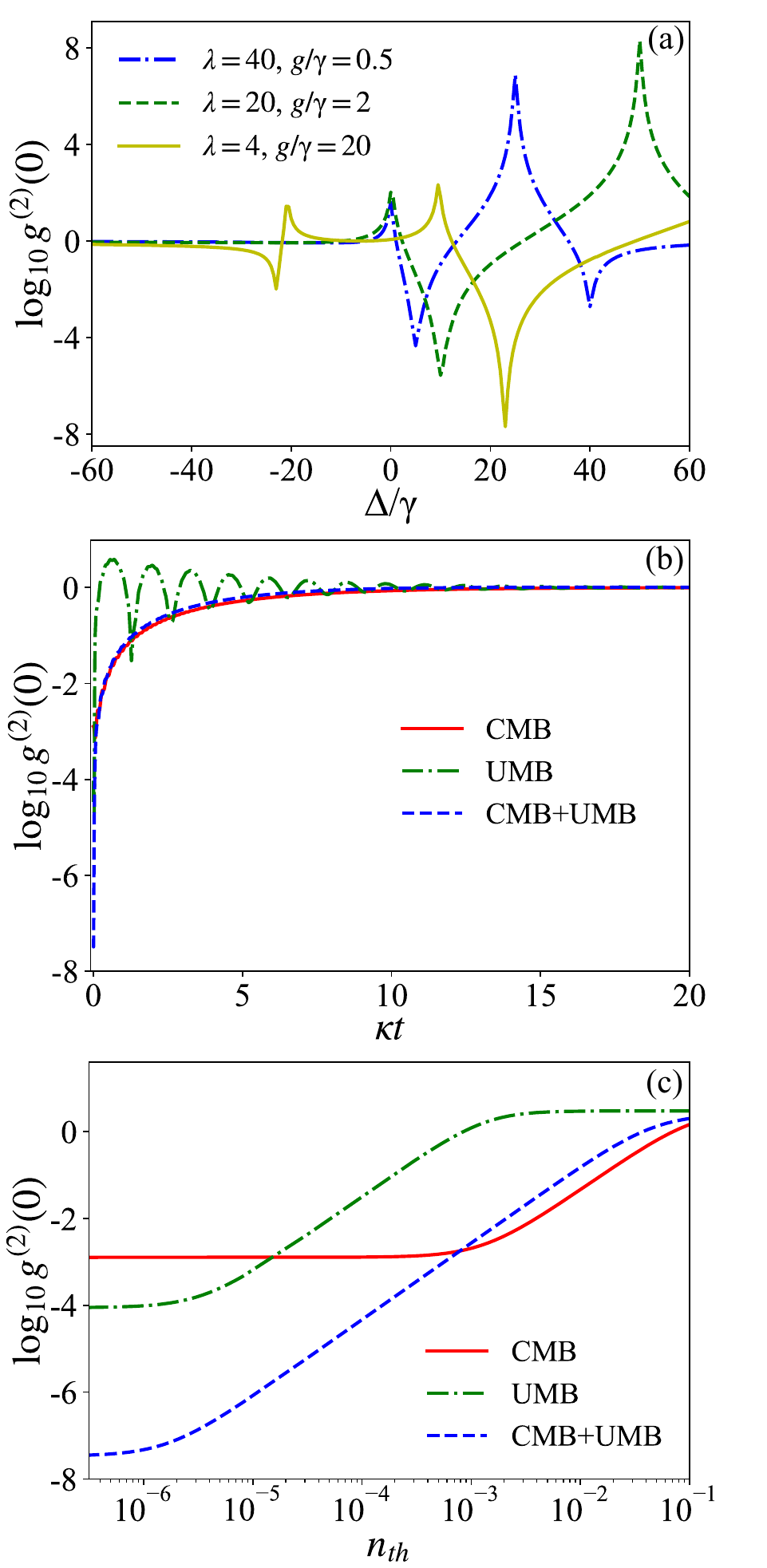}
\caption{(a) Appearance of the CMB and the UMB at the same time under different $\lambda$ and $g$ for (i) $\Delta_F/\gamma=5$, (ii) $\Delta_F/\gamma=10$ and (iii) $\Delta_F/\gamma=11.2$. (b) Time evolution of the second-order correlation function $\log_{10}g^{\left( 2 \right)}\left( t \right)$ for (i) CMB with $\Delta_F/\gamma=0$ and $\Delta/\gamma=20$, (ii) UMB with $\Delta_F/\gamma=0$ and $\Delta/\gamma=24.8$ and (iii) CMB+UMB with $\Delta_F/\gamma=11.2$ and $\Delta/\gamma=22.8$. (c) Correlation function $\log_{10}g^{\left( 2 \right)}\left( 0 \right)$ versus the thermal magnon occupation number $n_{th}$ for CMB, UMB, and CMB+UMB as the same as (b). The other parameters are the same as those in Fig.~\ref{fig2}.
\label{fig8}}
\end{figure}

Next, we discuss how to achieve both the CMB and the UMB in the presence of the downward shift of the UMB.
Combining Eq.~(\ref{eq:11}) and Eq.~(\ref{eq:15}), we can derive three intersection points of the CMB and the UMB:
\begin{align}
\Delta_F(1)= & -\frac{g \left( \lambda ^2-1 \right)}{2\lambda},\nonumber\\
\Delta _F(2)= & -\frac{g \left( \lambda +3 \sqrt{\lambda ^2-8} \right)}{8},\nonumber\\
\Delta _F(3)= &-\frac{g \left( \lambda -3 \sqrt{\lambda ^2-8} \right)}{8}.
\label{eq:21}
\end{align}
Here, the intersection points $\Delta _F(2)$ and $\Delta _F(3)$ can exist only when $\lambda>2\sqrt{2}$.
To verify intersection points of the CMB and the UMB, we plot the equal-time
second-order magnon correlation function $\log_{10} g^{\left( 2 \right)}\left( 0 \right)$
as a function of $\Delta$ and $\Delta_F$ with different values of $\lambda$ in Fig.~\ref{fig7}.
As shown in Fig.~\ref{fig7}(a), when $\lambda=1$, only one intersection point $\Delta _F(1)$ exists at $\Delta_F=0$ and $\Delta=g$.
When the parameter $\lambda>2\sqrt{2}$ is satisfied,
we can see three intersection points of the CMB and the UMB.
The intersection point $\Delta _F(2)$ is located between the intersection point $\Delta _F(1)$ and the intersection point $\Delta _F(3)$,
as shown in Figs.~\ref{fig7}(b) and (c).
Additionally, the intersection points $\Delta _F(1)$ and $\Delta _F(2)$ are close to the range of MIT's occurrence, leading to a weak MB.
In contrast, the intersection point $\Delta _F(3)$ moves away from the MIT
with the increase of the parameter $\lambda$, resulting in a very strong MB.

To further study the effect of the driving detuning $\Delta_F$ and the parameter $\lambda$ on the MB caused by the intersection point $\Delta _F(3)$, we present the $\log_{10}g^{\left( 2 \right)}\left( 0 \right)$ as a function of $\Delta$ for three cases in Fig.~\ref{fig8}(a). Even under the weak-coupling condition, the UMB and the CMB can occur simultaneously. As the coupling strength $g$ increases, the MB becomes stronger. When $\lambda=4$ and $g/\gamma=20$, the minimum value of the equal-time second-order correlation function $g^{\left( 2 \right)}\left( 0 \right)$ is approximately equal to $10^{-8}$, which is much lower than the values obtained from the single CMB or the single UMB.

To explore the dynamic features of the MB caused by $\Delta_F(3)$,
we calculate the time-delayed second-order correlation function by $g^{\left( 2 \right)}\left( t \right)={\left< m^{\dag}m^{\dag}(t)m(t)m \right>}/{\left< m^{\dag}m \right> ^2}$. The $\log_{10}g^{\left( 2 \right)}\left( t \right)$ with $\lambda=4$ and $g/\gamma=20$
is presented for three cases in Fig.~\ref{fig8}(b): (i) CMB with $\Delta_F=0$ and $\Delta/\gamma=-20$; (ii) UMB with $\Delta_F=0$ and $\Delta/\gamma=24.8$; and (iii) CMB+UMB with $\Delta_F=\Delta_F(3)$ and $\Delta/\gamma=22.8$. Here, $\Delta$ can be calculated using Eq.~(\ref{eq:12}) for the CMB and Eq.~(\ref{eq:15}) for the UMB, and $\Delta_F(3)$ can be obtained from Eq.~(\ref{eq:21}). As predicted,
$g^{\left( 2 \right)}\left( 0 \right)\ll 1$ and $g^{\left( 2 \right)}\left( 0 \right)<g^{\left( 2 \right)}\left( t \right)$ are obtained, which indicates the achievement of a strong MB. For the CMB, $g^{\left( 2 \right)}\left( t \right)$ gradually approaches 1, meaning that high time resolution is not necessary and the joint detection of single magnons becomes easier. For the UMB, $g^{\left( 2 \right)}\left( t \right)$ exhibits obvious oscillations and is greater than 1 at certain times, implying that high time resolution is required for observation. For the CMB+UMB, $g^{\left( 2 \right)}\left( t \right)$ smoothly approaches 0, similar to the CMB, and the value of $g^{\left( 2 \right)}\left( 0 \right)$ is less than that of either the CMB or the UMB. In other words, the CMB+UMB combines the advantages of both the UMB and the CMB while avoiding their respective disadvantages, thus representing an optimal scheme for generating the optimal MB.

Finally, let us investigate the influence of the thermal magnon occupation number $n_{th}$ on the equal-time second-order correlation function $g^{\left( 2 \right)}\left( 0 \right)$. In Fig.~\ref{fig8}(c), we plot the $\log_{10}g^{\left( 2 \right)}\left( 0 \right)$ as a function of $n_{th}$ for the CMB, the UMB and the CMB+UMB, with parameters consistent with those in Fig.~\ref{fig8}(b). We find that the equal-time second-order correlation function $g^{\left( 2 \right)}\left( 0 \right)$ gradually increases with the increase of $n_{th}$.
For the UMB, when $n_{th}=10^{-3}$, $g^{\left( 2 \right)}\left( 0 \right)$ is approximately equal to 1, which means that the anti-bunching effect disappears completely. However, only when $n_{th}=0.1$, $g^{\left( 2 \right)}\left( 0 \right)$ is approximately equal to 1 for the CMB and the CMB+UMB, and thus are quite robust against the thermal magnon occupation number. In experiments, the magnon has a high resonance frequency, so the thermal occupation of the magnon can be neglected. As a result, the thermal bath of the magnon has a negligible effect on our results.

\section{CONCLUSIONS}\label{4}
Based on the direct coupling of the magnon to the spin qubit of NV center, we study the physical mechanism for generating an optimal MB.
Different from previous schemes that strictly limit the coupling strength to achieve the MB, we show that when only the magnon is driven, the UMB can appear under both strong and weak coupling by adjusting the frequency detuning between the magnon and the spin qubit. Besides, when both the magnon and the spin qubit are driven, the CMB and the UMB can occur simultaneously under both strong and weak coupling by adjusting the frequency detuning. As a result, the equal-time correlation function can reach $10^{-8}$ and the time-delayed correlation function can avoid oscillations. This work provides a method to enhance the MB and relax the
requirements for coupling strength via frequency detuning, which can be used to generate a single-magnon source with high purity.
\begin{acknowledgments}
	P.-B. L. is supported by the National Natural Science Foundation of China under Grants No. W2411002 and No. 12375018.
	X.-F. P. is supported by the National Natural Science Foundation of China under Grants No. 124B2091.
\end{acknowledgments}

\appendix
\section{Analytical solutions for driving the magnon}\label{A1}
In this subsection, we show how to derive the equal-time second-order correlation function when the magnon is driven. In the weak driving field limit, This condition $C_{0g}\approx 1 \gg C_{0e},~ C_{1g}\gg C_{1e},~C_{2g}$ is valid. So the terms from $\left| 1,e \right>$ and $\left| 2,g \right>$ to $\left| 0,e \right>$ and $\left| 1,g \right>$ by driving field can be neglected. Substituting the wave function $\left| \psi \right> =C_{0g}\left| 0,g \right> +C_{0e}\left| 0,e \right> +C_{1g}\left| 1,g \right> +C_{1e}\left| 1,e \right> +C_{2g}\left| 2,g \right> $ and non-Hermitian Hamiltonian $H_\text{non}=H_\text{eff}-i\kappa( m^{\dag}m+\sigma _+\sigma _- )/2$ into the Schrödinger equation $i\partial \left| \psi \right> /\partial t=H_\text{non}\left| \psi \right>$, consequently, we have
\begin{align}
&i\dot{C}_{0e}=\left( \Delta -\Delta _F-i\kappa /2 \right) C_{0e}+gC_{1g},\nonumber\\
&i\dot{C}_{1g}=\Omega _mC_{0g}+gC_{0e}+\left( \Delta +\Delta _F-i\kappa /2 \right) C_{1g},\nonumber\\
&i\dot{C}_{1e}=\Omega _mC_{0e}+2\left( \Delta -i\kappa /2 \right) C_{1e}+\sqrt{2}gC_{2g},\nonumber\\
&i\dot{C}_{2g}=\sqrt{2}\Omega _mC_{1g}+\sqrt{2}gC_{1e}+2\left( \Delta +\Delta _F-i\kappa /2 \right) C_{2g}.\nonumber\\
\label{eqA1}
\end{align}
By setting the time derivatives in Eq. (\ref{eqA1}) to zero, we can obtain the steady-state solution
\begin{align}
&0=\left( \Delta -\Delta _F-i\kappa /2 \right) C_{0e}+gC_{1g},\nonumber\\
&0=\Omega _mC_{0g}+gC_{0e}+\left( \Delta +\Delta _F-i\kappa /2 \right) C_{1g},\nonumber\\
&0=\Omega _mC_{0e}+2\left( \Delta -i\kappa /2 \right) C_{1e}+\sqrt{2}gC_{2g},\nonumber\\
&0=\sqrt{2}\Omega _mC_{1g}+\sqrt{2}gC_{1e}+2\left( \Delta +\Delta _F-i\kappa /2 \right)C_{2g}.\nonumber\\
\label{eqA2}
\end{align}
From Eq. (\ref{eqA2}), we can get
\begin{align}
&C_{0e}=\frac{ -\Omega _mg }{C},\nonumber\\
&C_{1g}=\frac{ \Omega _m\left( \bar{\Delta} -\Delta _F \right)}{C},\nonumber\\
&C_{1e}=-\frac{-4\Omega _m^2g\bar{\Delta}}{2CD},\nonumber\\
&C_{2g}=\frac{-2\sqrt{2}\Omega _m^2\bar{\Delta}^2+2\sqrt{2}\Omega _m^2\Delta _F\bar{\Delta} -\sqrt{2} \Omega _m^2 g^2}{2CD}, \label{A3}
\end{align}
where
    \begin{align}
&\bar{\Delta}=\Delta -i\kappa /2,\nonumber\\
 &C=g^2-\bar{\Delta}^2+\Delta _{F}^{2},\nonumber\\
 &D=\bar{\Delta}^2+2\bar{\Delta}\Delta _F-g^2.
    \end{align}
Then, the equal-time correlation function $g^{\left( 2 \right)}\left( 0 \right)$ defined in Eq.~(\ref{eq:7}) can be obtained as
\begin{align}
g^{\left( 2 \right)}\left( 0 \right)&=\frac{2\left| C_{2g} \right|^2}{\left( \left| C_{1g} \right|^2+\left| C_{1e} \right|^2+2\left| C_{2g}^{\prime} \right|^2 \right) ^2}\nonumber\\
&\approx \frac{2\left| C_{2g} \right|^2}{\left| C_{1g} \right|^4}
=\frac{\left| B_1 \right|^2| C |^2 }{\left| A_1 \right|^2| D|^2},
\end{align}
where
\begin{align}
|C_{1g}|^2&=\frac{\Omega _{m}^{2}\left| A_1 \right|}{\left| C \right|^2},\quad
|C_{2g}|^2=\frac{\Omega _{m}^{4}\left| B_1 \right|^2}{2\left| C \right|^2\left| D \right|^2},\nonumber\\
\left| A_1 \right|^2&=\left[ \left( \Delta -\Delta _F \right) ^2+\kappa ^2/4 \right] ^2,\nonumber\\
\left| B_1 \right|^2&=\left( -2\Delta ^2+2\Delta _F\Delta -g^2+\kappa ^2/2 \right) ^2+\kappa ^2\left( 2\Delta -\Delta _F \right) ^2.\nonumber\\
\left| C \right|^2&=\left( g^2+\Delta _{F}^{2}-\Delta ^2+\kappa ^2/4 \right) ^2+\kappa ^2\Delta ^2,\nonumber\\
\left| D \right|^2&=\left( 2\Delta ^2+2\Delta _F\Delta -g^2-\kappa ^2/2 \right) ^2+\kappa ^2\left( 2\Delta +\Delta _F \right) ^2\nonumber\
\end{align}

\section{Analytical solutions for driving the magnon and the spin qubit}\label{B1}
In this subsection, we derive the equal-time second-order correlation function when the magnon and the spin qubit are both driven. Substituting the wave function $\left| \psi^{\prime} \right> =C_{0g}^{\prime}\left| 0,g \right> +C_{0e}^{\prime}\left| 0,e \right> +C_{1g}^{\prime}\left| 1,g \right> +C_{1e}^{\prime}\left| 1,e \right> +C_{2g}^{\prime}\left| 2,g \right> $ and non-Hermitian Hamiltonian $H_\text{non}^{\prime}=H_\text{eff}^{\prime}-i\kappa( m^{\dag}m+\sigma _+\sigma _- )/2$ into the Schrödinger equation $i\partial \left| \psi^{\prime} \right> /\partial t=H_\text{non}^{\prime}\left| \psi^{\prime} \right>$, consequently, we have
\begin{align}
&i\dot{C}_{0e}^{\prime}=\Omega _\text{NV}C_{0g}^{\prime}+\left( \Delta -\Delta _F-i\kappa /2 \right) C_{0e}^{\prime}+gC_{1g}^{\prime},\nonumber\\
&i\dot{C}_{1g}^{\prime}=\Omega _mC_{0g}^{\prime}+gC_{0e}^{\prime}+\left( \Delta +\Delta _F-i\kappa /2 \right) C_{1g}^{\prime},\nonumber\\
&i\dot{C}_{1e}^{\prime}=\Omega _mC_{0e}^{\prime}+\Omega _\text{NV}C_{1g}^{\prime}+2\left( \Delta -i\kappa /2 \right) C_{1e}^{\prime}+\sqrt{2}gC_{2g}^{\prime},\nonumber\\
&i\dot{C}_{2g}^{\prime}=\sqrt{2}\Omega _mC_{1g}^{\prime}+\sqrt{2}gC_{1e}^{\prime}+2\left( \Delta +\Delta _F-i\kappa /2 \right) C_{2g}^{\prime}.\nonumber\\
\end{align}
Then steady-state solution can be obtained as
\begin{align}
&0=\Omega _\text{NV}C_{0g}^{\prime}+\left( \Delta -\Delta _F-i\kappa /2 \right) C_{0e}^{\prime}+gC_{1g}^{\prime},\nonumber\\
&0=\Omega _mC_{0g}^{\prime}+gC_{0e}^{\prime}+\left( \Delta +\Delta _F-i\kappa /2 \right) C_{1g}^{\prime},\nonumber\\
&0=\Omega _mC_{0e}^{\prime}+\Omega _\text{NV}C_{1g}^{\prime}+2\left( \Delta -i\kappa /2 \right) C_{1e}^{\prime}+\sqrt{2}gC_{2g}^{\prime},\nonumber\\
&0=\sqrt{2}\Omega _mC_{1g}^{\prime}+\sqrt{2}gC_{1e}^{\prime}+2\left( \Delta +\Delta _F-i\kappa /2 \right)C_{2g}^{\prime}.\nonumber\\
\label{eqb2}
\end{align}
From Eq. (\ref{eqb2}), we have
\begin{widetext}
\begin{align}
&C_{0e}^{\prime}=\frac{\left( \Omega _\text{NV}\left( \bar{\Delta} +\Delta _F \right) -\Omega _mg \right)}{C},\nonumber\\
&C_{1g}^{\prime}=\frac{\left( \Omega _m\left( \bar{\Delta} -\Delta _F \right) -\Omega _\text{NV}g \right)}{C},\nonumber\\
&C_{1e}^{\prime}=-\frac{4\Omega _m\Omega _\text{NV}\bar{\Delta}^2+\left( 4\Omega _m\Omega _\text{NV}\Delta _F-4\Omega _m^2g-2\Omega _\text{NV}^2g \right) \bar{\Delta} +2\Omega _m\Omega _\text{NV}g^2-2\Omega _\text{NV}^2g\Delta _F}{2CD},\nonumber\\
&C_{2g}^{\prime}=\frac{-2\sqrt{2}\Omega _m^2\bar{\Delta}^2+\left( 4\sqrt{2}\Omega _m\Omega _\text{NV}g+2\sqrt{2}\Omega _m^2\Delta _F \right) \bar{\Delta} -\sqrt{2}\left( \Omega _m^2+\Omega _\text{NV}^2 \right) g^2}{2CD}, \label{B3}
\end{align}
\end{widetext}
where
\begin{align}
&\bar{\Delta}=\Delta -i\kappa /2,\nonumber\\
 &C=g^2-\bar{\Delta}^2+\Delta _{F}^{2},\nonumber\\
 &D=\bar{\Delta}^2+2\bar{\Delta}\Delta _F-g^2.
\end{align}
Then, the correlation function $g^{\left( 2 \right)}\left( 0 \right)$ defined in Eq.~(\ref{eq:7}) can be written as
\begin{align}
g^{\left( 2 \right)}\left( 0 \right)&=\frac{2\left| C_{2g}^{\prime} \right|^2}{\left( \left| C_{1g}^{\prime} \right|^2+\left| C_{1e}^{\prime} \right|^2+2\left| C_{2g}^{\prime} \right|^2 \right) ^2}\nonumber\\
&\approx \frac{2\left| C_{2g}^{\prime} \right|^2}{\left| C_{1g}^{\prime} \right|^4}
=\frac{\left| B_2 \right|^2| C |^2 }{\left| A_2 \right|^2| D|^2},
\end{align}
where
\begin{align}
|C_{1g}^{\prime}|^2&=\frac{\Omega _{m}^{2}\left| A_2 \right|}{\left| C \right|^2},\quad
|C_{2g}^{\prime}|^2=\frac{\Omega _{m}^{4}\left| B_2 \right|^2}{2\left| C \right|^2\left| D \right|^2},\nonumber\\
\left| A_2 \right|^2&=\left[ \left( \Delta -\Delta _F-\lambda g \right) ^2+\kappa ^2/4 \right] ^2,\nonumber\\
\left| B_2 \right|^2&=\left[ -2\left( \Delta ^2-\kappa ^2/4 \right) +\left( 4\lambda g+2\Delta _F \right) \Delta -\left( 1+\lambda ^2 \right) g^2 \right] ^2\nonumber\\
&+\kappa ^2\left[ 2\Delta -2\lambda g-\Delta _F \right] ^2.\nonumber\\
\left| C \right|^2&=\left( g^2+\Delta _{F}^{2}-\Delta ^2+\kappa ^2/4 \right) ^2+\kappa ^2\Delta ^2,\nonumber\\
\left| D \right|^2&=\left( 2\Delta ^2+2\Delta _F\Delta -g^2-\kappa ^2/2 \right) ^2+\kappa ^2\left( 2\Delta +\Delta _F \right) ^2\nonumber\
\end{align}
Here, $\lambda=\Omega_\text{NV}/{\Omega_m}$ is the ratio of the Rabi frequency of the qubit to that of the magnon.


\begin{thebibliography}{66}%
	\makeatletter
	\providecommand \@ifxundefined [1]{%
	 \@ifx{#1\undefined}
	}%
	\providecommand \@ifnum [1]{%
	 \ifnum #1\expandafter \@firstoftwo
	 \else \expandafter \@secondoftwo
	 \fi
	}%
	\providecommand \@ifx [1]{%
	 \ifx #1\expandafter \@firstoftwo
	 \else \expandafter \@secondoftwo
	 \fi
	}%
	\providecommand \natexlab [1]{#1}%
	\providecommand \enquote  [1]{``#1''}%
	\providecommand \bibnamefont  [1]{#1}%
	\providecommand \bibfnamefont [1]{#1}%
	\providecommand \citenamefont [1]{#1}%
	\providecommand \href@noop [0]{\@secondoftwo}%
	\providecommand \href [0]{\begingroup \@sanitize@url \@href}%
	\providecommand \@href[1]{\@@startlink{#1}\@@href}%
	\providecommand \@@href[1]{\endgroup#1\@@endlink}%
	\providecommand \@sanitize@url [0]{\catcode `\\12\catcode `\$12\catcode `\&12\catcode `\#12\catcode `\^12\catcode `\_12\catcode `\%12\relax}%
	\providecommand \@@startlink[1]{}%
	\providecommand \@@endlink[0]{}%
	\providecommand \url  [0]{\begingroup\@sanitize@url \@url }%
	\providecommand \@url [1]{\endgroup\@href {#1}{\urlprefix }}%
	\providecommand \urlprefix  [0]{URL }%
	\providecommand \Eprint [0]{\href }%
	\providecommand \doibase [0]{http://dx.doi.org/}%
	\providecommand \selectlanguage [0]{\@gobble}%
	\providecommand \bibinfo  [0]{\@secondoftwo}%
	\providecommand \bibfield  [0]{\@secondoftwo}%
	\providecommand \translation [1]{[#1]}%
	\providecommand \BibitemOpen [0]{}%
	\providecommand \bibitemStop [0]{}%
	\providecommand \bibitemNoStop [0]{.\EOS\space}%
	\providecommand \EOS [0]{\spacefactor3000\relax}%
	\providecommand \BibitemShut  [1]{\csname bibitem#1\endcsname}%
	\let\auto@bib@innerbib\@empty
	\bibitem [{\citenamefont {Knill}\ \  {et~al.}(2001)\citenamefont {Knill}, \citenamefont {Laflamme},\ and\ \citenamefont {Milburn}}]{knill2001scheme}%
	  \BibitemOpen
	  \bibfield  {author} {\bibinfo {author} {\bibfnamefont {E.}~\bibnamefont {Knill}}, \bibinfo {author} {\bibfnamefont {R.}~\bibnamefont {Laflamme}}, \ and\ \bibinfo {author} {\bibfnamefont {G.~J.}\ \bibnamefont {Milburn}},\ }\bibfield  {title} {\  {\bibinfo {title} {A scheme for efficient quantum computation with linear optics},\ }}\href {https://doi.org/10.1038/35051009} {\bibfield  {journal} {\bibinfo  {journal} {Nature}\ }\textbf {\bibinfo {volume} {409}},\ \bibinfo {pages} {46} (\bibinfo {year} {2001})}\BibitemShut {NoStop}%
	\bibitem [{\citenamefont {Kok}\ \  {et~al.}(2007)\citenamefont {Kok}, \citenamefont {Munro}, \citenamefont {Nemoto}, \citenamefont {Ralph}, \citenamefont {Dowling},\ and\ \citenamefont {Milburn}}]{kok2007linear}%
	  \BibitemOpen
	  \bibfield  {author} {\bibinfo {author} {\bibfnamefont {P.}~\bibnamefont {Kok}}, \bibinfo {author} {\bibfnamefont {W.~J.}\ \bibnamefont {Munro}}, \bibinfo {author} {\bibfnamefont {K.}~\bibnamefont {Nemoto}}, \bibinfo {author} {\bibfnamefont {T.~C.}\ \bibnamefont {Ralph}}, \bibinfo {author} {\bibfnamefont {J.~P.}\ \bibnamefont {Dowling}}, \ and\ \bibinfo {author} {\bibfnamefont {G.~J.}\ \bibnamefont {Milburn}},\ }\bibfield  {title} {\  {\bibinfo {title} {Linear optical quantum computing with photonic qubits},\ }}\href {\doibase 10.1103/RevModPhys.79.135} {\bibfield  {journal} {\bibinfo  {journal} {Rev. Mod. Phys.}\ }\textbf {\bibinfo {volume} {79}},\ \bibinfo {pages} {135} (\bibinfo {year} {2007})}\BibitemShut {NoStop}%
	\bibitem [{\citenamefont {Kimble}(2008)}]{kimble2008quantum}%
	  \BibitemOpen
	  \bibfield  {author} {\bibinfo {author} {\bibfnamefont {H.~J.}\ \bibnamefont {Kimble}},\ }\bibfield  {title} {\  {\bibinfo {title} {The quantum internet},\ }}\href {https://doi.org/10.1038/nature07127} {\bibfield  {journal} {\bibinfo  {journal} {Nature}\ }\textbf {\bibinfo {volume} {453}},\ \bibinfo {pages} {1023} (\bibinfo {year} {2008})}\BibitemShut {NoStop}%
	\bibitem [{\citenamefont {Gilleo}\ and\ \citenamefont {Geller}(1958)}]{gilleo1958magnetic}%
	  \BibitemOpen
	  \bibfield  {author} {\bibinfo {author} {\bibfnamefont {M.~A.}\ \bibnamefont {Gilleo}}\ and\ \bibinfo {author} {\bibfnamefont {S.}~\bibnamefont {Geller}},\ }\bibfield  {title} {\  {\bibinfo {title} {Magnetic and crystallographic properties of substituted yttrium-iron garnet, $3{\mathrm{y}}_{2}{\mathrm{o}}_{3}\ifmmode\cdot\else\textperiodcentered\fi{}x{\mathrm{m}}_{2}{\mathrm{o}}_{3}\ifmmode\cdot\else\textperiodcentered\fi{}(5\ensuremath{-}x){\mathrm{fe}}_{2}{\mathrm{o}}_{3}$},\ }}\href {\doibase 10.1103/PhysRev.110.73} {\bibfield  {journal} {\bibinfo  {journal} {Phys. Rev.}\ }\textbf {\bibinfo {volume} {110}},\ \bibinfo {pages} {73} (\bibinfo {year} {1958})}\BibitemShut {NoStop}%
	\bibitem [{\citenamefont {Tabuchi}\ \  {et~al.}(2014)\citenamefont {Tabuchi}, \citenamefont {Ishino}, \citenamefont {Ishikawa}, \citenamefont {Yamazaki}, \citenamefont {Usami},\ and\ \citenamefont {Nakamura}}]{PhysRevLett.113.083603}%
	  \BibitemOpen
	  \bibfield  {author} {\bibinfo {author} {\bibfnamefont {Y.}~\bibnamefont {Tabuchi}}, \bibinfo {author} {\bibfnamefont {S.}~\bibnamefont {Ishino}}, \bibinfo {author} {\bibfnamefont {T.}~\bibnamefont {Ishikawa}}, \bibinfo {author} {\bibfnamefont {R.}~\bibnamefont {Yamazaki}}, \bibinfo {author} {\bibfnamefont {K.}~\bibnamefont {Usami}}, \ and\ \bibinfo {author} {\bibfnamefont {Y.}~\bibnamefont {Nakamura}},\ }\bibfield  {title} {\  {\bibinfo {title} {Hybridizing ferromagnetic magnons and microwave photons in the quantum limit},\ }}\href {\doibase 10.1103/PhysRevLett.113.083603} {\bibfield  {journal} {\bibinfo  {journal} {Phys. Rev. Lett.}\ }\textbf {\bibinfo {volume} {113}},\ \bibinfo {pages} {083603} (\bibinfo {year} {2014})}\BibitemShut {NoStop}%
	\bibitem [{\citenamefont {Barker}\ and\ \citenamefont {Bauer}(2016)}]{barker2016thermal}%
	  \BibitemOpen
	  \bibfield  {author} {\bibinfo {author} {\bibfnamefont {J.}~\bibnamefont {Barker}}\ and\ \bibinfo {author} {\bibfnamefont {G.~E.~W.}\ \bibnamefont {Bauer}},\ }\bibfield  {title} {\  {\bibinfo {title} {Thermal spin dynamics of yttrium iron garnet},\ }}\href {\doibase 10.1103/PhysRevLett.117.217201} {\bibfield  {journal} {\bibinfo  {journal} {Phys. Rev. Lett.}\ }\textbf {\bibinfo {volume} {117}},\ \bibinfo {pages} {217201} (\bibinfo {year} {2016})}\BibitemShut {NoStop}%
	\bibitem [{\citenamefont {Collet}\ \  {et~al.}(2016)\citenamefont {Collet}, \citenamefont {De~Milly}, \citenamefont {d’Allivy Kelly}, \citenamefont {Naletov}, \citenamefont {Bernard}, \citenamefont {Bortolotti}, \citenamefont {Ben~Youssef}, \citenamefont {Demidov}, \citenamefont {Demokritov}, \citenamefont {Prieto} \  {et~al.}}]{collet2016generation}%
	  \BibitemOpen
	  \bibfield  {author} {\bibinfo {author} {\bibfnamefont {M.}~\bibnamefont {Collet}}, \bibinfo {author} {\bibfnamefont {X.}~\bibnamefont {De~Milly}}, \bibinfo {author} {\bibfnamefont {O.}~\bibnamefont {d’Allivy Kelly}}, \bibinfo {author} {\bibfnamefont {V.~V.}\ \bibnamefont {Naletov}}, \bibinfo {author} {\bibfnamefont {R.}~\bibnamefont {Bernard}}, \bibinfo {author} {\bibfnamefont {P.}~\bibnamefont {Bortolotti}}, \bibinfo {author} {\bibfnamefont {J.}~\bibnamefont {Ben~Youssef}}, \bibinfo {author} {\bibfnamefont {V.}~\bibnamefont {Demidov}}, \bibinfo {author} {\bibfnamefont {S.}~\bibnamefont {Demokritov}}, \bibinfo {author} {\bibfnamefont {J.~L.}\ \bibnamefont {Prieto}},  \  {et~al.},\ }\bibfield  {title} {\  {\bibinfo {title} {Generation of coherent spin-wave modes in yttrium iron garnet microdiscs by spin--orbit torque},\ }}\href {https://doi.org/10.1038/ncomms10377} {\bibfield  {journal} {\bibinfo  {journal} {Nat. Commun.}\ }\textbf {\bibinfo {volume} {7}},\ \bibinfo {pages} {10377} (\bibinfo {year} {2016})}\BibitemShut {NoStop}%
	\bibitem [{\citenamefont {Princep}\ \  {et~al.}(2017)\citenamefont {Princep}, \citenamefont {Ewings}, \citenamefont {Ward}, \citenamefont {T{\'o}th}, \citenamefont {Dubs}, \citenamefont {Prabhakaran},\ and\ \citenamefont {Boothroyd}}]{princep2017full}%
	  \BibitemOpen
	  \bibfield  {author} {\bibinfo {author} {\bibfnamefont {A.~J.}\ \bibnamefont {Princep}}, \bibinfo {author} {\bibfnamefont {R.~A.}\ \bibnamefont {Ewings}}, \bibinfo {author} {\bibfnamefont {S.}~\bibnamefont {Ward}}, \bibinfo {author} {\bibfnamefont {S.}~\bibnamefont {T{\'o}th}}, \bibinfo {author} {\bibfnamefont {C.}~\bibnamefont {Dubs}}, \bibinfo {author} {\bibfnamefont {D.}~\bibnamefont {Prabhakaran}}, \ and\ \bibinfo {author} {\bibfnamefont {A.~T.}\ \bibnamefont {Boothroyd}},\ }\bibfield  {title} {\  {\bibinfo {title} {The full magnon spectrum of yttrium iron garnet},\ }}\href {https://doi.org/10.1038/s41535-017-0067-y} {\bibfield  {journal} {\bibinfo  {journal} {npj Quant. Mater.}\ }\textbf {\bibinfo {volume} {2}},\ \bibinfo {pages} {63} (\bibinfo {year} {2017})}\BibitemShut {NoStop}%
	\bibitem [{\citenamefont {Wei}\ \  {et~al.}(2022)\citenamefont {Wei}, \citenamefont {Santos}, \citenamefont {Lusero}, \citenamefont {Bauer}, \citenamefont {Ben~Youssef},\ and\ \citenamefont {van Wees}}]{wei2022giant}%
	  \BibitemOpen
	  \bibfield  {author} {\bibinfo {author} {\bibfnamefont {X.-Y.}\ \bibnamefont {Wei}}, \bibinfo {author} {\bibfnamefont {O.~A.}\ \bibnamefont {Santos}}, \bibinfo {author} {\bibfnamefont {C.~S.}\ \bibnamefont {Lusero}}, \bibinfo {author} {\bibfnamefont {G.}~\bibnamefont {Bauer}}, \bibinfo {author} {\bibfnamefont {J.}~\bibnamefont {Ben~Youssef}}, \ and\ \bibinfo {author} {\bibfnamefont {B.}~\bibnamefont {van Wees}},\ }\bibfield  {title} {\  {\bibinfo {title} {Giant magnon spin conductivity in ultrathin yttrium iron garnet films},\ }}\href {https://doi.org/10.1038/s41563-022-01369-0} {\bibfield  {journal} {\bibinfo  {journal} {Nat. Mater.}\ }\textbf {\bibinfo {volume} {21}},\ \bibinfo {pages} {1352} (\bibinfo {year} {2022})}\BibitemShut {NoStop}%
	\bibitem [{\citenamefont {Zhang}\ \  {et~al.}(2016)\citenamefont {Zhang}, \citenamefont {Zou}, \citenamefont {Jiang},\ and\ \citenamefont {Tang}}]{zhang2016cavity}%
	  \BibitemOpen
	  \bibfield  {author} {\bibinfo {author} {\bibfnamefont {X.}~\bibnamefont {Zhang}}, \bibinfo {author} {\bibfnamefont {C.-L.}\ \bibnamefont {Zou}}, \bibinfo {author} {\bibfnamefont {L.}~\bibnamefont {Jiang}}, \ and\ \bibinfo {author} {\bibfnamefont {H.~X.}\ \bibnamefont {Tang}},\ }\bibfield  {title} {\  {\bibinfo {title} {Cavity magnomechanics},\ }}\href {\doibase 10.1126/sciadv.1501286} {\bibfield  {journal} {\bibinfo  {journal} {Science Advances}\ }\textbf {\bibinfo {volume} {2}},\ \bibinfo {pages} {e1501286} (\bibinfo {year} {2016})}\BibitemShut {NoStop}%
	\bibitem [{\citenamefont {Zhang}\ \  {et~al.}(2014)\citenamefont {Zhang}, \citenamefont {Zou}, \citenamefont {Jiang},\ and\ \citenamefont {Tang}}]{zhang2014strongly}%
	  \BibitemOpen
	  \bibfield  {author} {\bibinfo {author} {\bibfnamefont {X.}~\bibnamefont {Zhang}}, \bibinfo {author} {\bibfnamefont {C.-L.}\ \bibnamefont {Zou}}, \bibinfo {author} {\bibfnamefont {L.}~\bibnamefont {Jiang}}, \ and\ \bibinfo {author} {\bibfnamefont {H.~X.}\ \bibnamefont {Tang}},\ }\bibfield  {title} {\  {\bibinfo {title} {Strongly coupled magnons and cavity microwave photons},\ }}\href {\doibase 10.1103/PhysRevLett.113.156401} {\bibfield  {journal} {\bibinfo  {journal} {Phys. Rev. Lett.}\ }\textbf {\bibinfo {volume} {113}},\ \bibinfo {pages} {156401} (\bibinfo {year} {2014})}\BibitemShut {NoStop}%
	\bibitem [{\citenamefont {Huebl}\ \  {et~al.}(2013)\citenamefont {Huebl}, \citenamefont {Zollitsch}, \citenamefont {Lotze}, \citenamefont {Hocke}, \citenamefont {Greifenstein}, \citenamefont {Marx}, \citenamefont {Gross},\ and\ \citenamefont {Goennenwein}}]{huebl2013high}%
	  \BibitemOpen
	  \bibfield  {author} {\bibinfo {author} {\bibfnamefont {H.}~\bibnamefont {Huebl}}, \bibinfo {author} {\bibfnamefont {C.~W.}\ \bibnamefont {Zollitsch}}, \bibinfo {author} {\bibfnamefont {J.}~\bibnamefont {Lotze}}, \bibinfo {author} {\bibfnamefont {F.}~\bibnamefont {Hocke}}, \bibinfo {author} {\bibfnamefont {M.}~\bibnamefont {Greifenstein}}, \bibinfo {author} {\bibfnamefont {A.}~\bibnamefont {Marx}}, \bibinfo {author} {\bibfnamefont {R.}~\bibnamefont {Gross}}, \ and\ \bibinfo {author} {\bibfnamefont {S.~T.~B.}\ \bibnamefont {Goennenwein}},\ }\bibfield  {title} {\  {\bibinfo {title} {High cooperativity in coupled microwave resonator ferrimagnetic insulator hybrids},\ }}\href {\doibase 10.1103/PhysRevLett.111.127003} {\bibfield  {journal} {\bibinfo  {journal} {Phys. Rev. Lett.}\ }\textbf {\bibinfo {volume} {111}},\ \bibinfo {pages} {127003} (\bibinfo {year} {2013})}\BibitemShut {NoStop}%
	\bibitem [{\citenamefont {Zare~Rameshti}\ \  {et~al.}(2015)\citenamefont {Zare~Rameshti}, \citenamefont {Cao},\ and\ \citenamefont {Bauer}}]{zare2015magnetic}%
	  \BibitemOpen
	  \bibfield  {author} {\bibinfo {author} {\bibfnamefont {B.}~\bibnamefont {Zare~Rameshti}}, \bibinfo {author} {\bibfnamefont {Y.}~\bibnamefont {Cao}}, \ and\ \bibinfo {author} {\bibfnamefont {G.~E.~W.}\ \bibnamefont {Bauer}},\ }\bibfield  {title} {\  {\bibinfo {title} {Magnetic spheres in microwave cavities},\ }}\href {\doibase 10.1103/PhysRevB.91.214430} {\bibfield  {journal} {\bibinfo  {journal} {Phys. Rev. B}\ }\textbf {\bibinfo {volume} {91}},\ \bibinfo {pages} {214430} (\bibinfo {year} {2015})}\BibitemShut {NoStop}%
	\bibitem [{\citenamefont {Maier-Flaig}\ \  {et~al.}(2016)\citenamefont {Maier-Flaig}, \citenamefont {Harder}, \citenamefont {Gross}, \citenamefont {Huebl},\ and\ \citenamefont {Goennenwein}}]{maier2016spin}%
	  \BibitemOpen
	  \bibfield  {author} {\bibinfo {author} {\bibfnamefont {H.}~\bibnamefont {Maier-Flaig}}, \bibinfo {author} {\bibfnamefont {M.}~\bibnamefont {Harder}}, \bibinfo {author} {\bibfnamefont {R.}~\bibnamefont {Gross}}, \bibinfo {author} {\bibfnamefont {H.}~\bibnamefont {Huebl}}, \ and\ \bibinfo {author} {\bibfnamefont {S.~T.~B.}\ \bibnamefont {Goennenwein}},\ }\bibfield  {title} {\  {\bibinfo {title} {Spin pumping in strongly coupled magnon-photon systems},\ }}\href {\doibase 10.1103/PhysRevB.94.054433} {\bibfield  {journal} {\bibinfo  {journal} {Phys. Rev. B}\ }\textbf {\bibinfo {volume} {94}},\ \bibinfo {pages} {054433} (\bibinfo {year} {2016})}\BibitemShut {NoStop}%
	\bibitem [{\citenamefont {Tabuchi}\ \  {et~al.}(2015)\citenamefont {Tabuchi}, \citenamefont {Ishino}, \citenamefont {Noguchi}, \citenamefont {Ishikawa}, \citenamefont {Yamazaki}, \citenamefont {Usami},\ and\ \citenamefont {Nakamura}}]{tabuchi2015coherent}%
	  \BibitemOpen
	  \bibfield  {author} {\bibinfo {author} {\bibfnamefont {Y.}~\bibnamefont {Tabuchi}}, \bibinfo {author} {\bibfnamefont {S.}~\bibnamefont {Ishino}}, \bibinfo {author} {\bibfnamefont {A.}~\bibnamefont {Noguchi}}, \bibinfo {author} {\bibfnamefont {T.}~\bibnamefont {Ishikawa}}, \bibinfo {author} {\bibfnamefont {R.}~\bibnamefont {Yamazaki}}, \bibinfo {author} {\bibfnamefont {K.}~\bibnamefont {Usami}}, \ and\ \bibinfo {author} {\bibfnamefont {Y.}~\bibnamefont {Nakamura}},\ }\bibfield  {title} {\  {\bibinfo {title} {Coherent coupling between a ferromagnetic magnon and a superconducting qubit},\ }}\href {\doibase 10.1126/science.aaa3693} {\bibfield  {journal} {\bibinfo  {journal} {Science}\ }\textbf {\bibinfo {volume} {349}},\ \bibinfo {pages} {405} (\bibinfo {year} {2015})}\BibitemShut {NoStop}%
	\bibitem [{\citenamefont {Li}\ and\ \citenamefont {Long}(2016)}]{li2016hyperparallel}%
	  \BibitemOpen
	  \bibfield  {author} {\bibinfo {author} {\bibfnamefont {T.}~\bibnamefont {Li}}\ and\ \bibinfo {author} {\bibfnamefont {G.-L.}\ \bibnamefont {Long}},\ }\bibfield  {title} {\  {\bibinfo {title} {Hyperparallel optical quantum computation assisted by atomic ensembles embedded in double-sided optical cavities},\ }}\href {\doibase 10.1103/PhysRevA.94.022343} {\bibfield  {journal} {\bibinfo  {journal} {Phys. Rev. A}\ }\textbf {\bibinfo {volume} {94}},\ \bibinfo {pages} {022343} (\bibinfo {year} {2016})}\BibitemShut {NoStop}%
	\bibitem [{\citenamefont {Clerk}\ \  {et~al.}(2020)\citenamefont {Clerk}, \citenamefont {Lehnert}, \citenamefont {Bertet}, \citenamefont {Petta},\ and\ \citenamefont {Nakamura}}]{clerk2020hybrid}%
	  \BibitemOpen
	  \bibfield  {author} {\bibinfo {author} {\bibfnamefont {A.}~\bibnamefont {Clerk}}, \bibinfo {author} {\bibfnamefont {K.}~\bibnamefont {Lehnert}}, \bibinfo {author} {\bibfnamefont {P.}~\bibnamefont {Bertet}}, \bibinfo {author} {\bibfnamefont {J.}~\bibnamefont {Petta}}, \ and\ \bibinfo {author} {\bibfnamefont {Y.}~\bibnamefont {Nakamura}},\ }\bibfield  {title} {\  {\bibinfo {title} {Hybrid quantum systems with circuit quantum electrodynamics},\ }}\href {https://doi.org/10.1038/s41567-020-0797-9} {\bibfield  {journal} {\bibinfo  {journal} {Nat. Phys.}\ }\textbf {\bibinfo {volume} {16}},\ \bibinfo {pages} {257} (\bibinfo {year} {2020})}\BibitemShut {NoStop}%
	\bibitem [{\citenamefont {Soykal}\ and\ \citenamefont {Flatt\'e}(2010)}]{soykal2010strong}%
	  \BibitemOpen
	  \bibfield  {author} {\bibinfo {author} {\bibfnamefont {O.~O.}\ \bibnamefont {Soykal}}\ and\ \bibinfo {author} {\bibfnamefont {M.~E.}\ \bibnamefont {Flatt\'e}},\ }\bibfield  {title} {\  {\bibinfo {title} {Strong field interactions between a nanomagnet and a photonic cavity},\ }}\href {\doibase 10.1103/PhysRevLett.104.077202} {\bibfield  {journal} {\bibinfo  {journal} {Phys. Rev. Lett.}\ }\textbf {\bibinfo {volume} {104}},\ \bibinfo {pages} {077202} (\bibinfo {year} {2010})}\BibitemShut {NoStop}%
	\bibitem [{\citenamefont {Bai}\ \  {et~al.}(2015)\citenamefont {Bai}, \citenamefont {Harder}, \citenamefont {Chen}, \citenamefont {Fan}, \citenamefont {Xiao},\ and\ \citenamefont {Hu}}]{bai2015spin}%
	  \BibitemOpen
	  \bibfield  {author} {\bibinfo {author} {\bibfnamefont {L.}~\bibnamefont {Bai}}, \bibinfo {author} {\bibfnamefont {M.}~\bibnamefont {Harder}}, \bibinfo {author} {\bibfnamefont {Y.~P.}\ \bibnamefont {Chen}}, \bibinfo {author} {\bibfnamefont {X.}~\bibnamefont {Fan}}, \bibinfo {author} {\bibfnamefont {J.~Q.}\ \bibnamefont {Xiao}}, \ and\ \bibinfo {author} {\bibfnamefont {C.-M.}\ \bibnamefont {Hu}},\ }\bibfield  {title} {\  {\bibinfo {title} {Spin pumping in electrodynamically coupled magnon-photon systems},\ }}\href {\doibase 10.1103/PhysRevLett.114.227201} {\bibfield  {journal} {\bibinfo  {journal} {Phys. Rev. Lett.}\ }\textbf {\bibinfo {volume} {114}},\ \bibinfo {pages} {227201} (\bibinfo {year} {2015})}\BibitemShut {NoStop}%
	\bibitem [{\citenamefont {Viola~Kusminskiy}\ \  {et~al.}(2016)\citenamefont {Viola~Kusminskiy}, \citenamefont {Tang},\ and\ \citenamefont {Marquardt}}]{viola2016coupled}%
	  \BibitemOpen
	  \bibfield  {author} {\bibinfo {author} {\bibfnamefont {S.}~\bibnamefont {Viola~Kusminskiy}}, \bibinfo {author} {\bibfnamefont {H.~X.}\ \bibnamefont {Tang}}, \ and\ \bibinfo {author} {\bibfnamefont {F.}~\bibnamefont {Marquardt}},\ }\bibfield  {title} {\  {\bibinfo {title} {Coupled spin-light dynamics in cavity optomagnonics},\ }}\href {\doibase 10.1103/PhysRevA.94.033821} {\bibfield  {journal} {\bibinfo  {journal} {Phys. Rev. A}\ }\textbf {\bibinfo {volume} {94}},\ \bibinfo {pages} {033821} (\bibinfo {year} {2016})}\BibitemShut {NoStop}%
	\bibitem [{\citenamefont {Haigh}\ \  {et~al.}(2016)\citenamefont {Haigh}, \citenamefont {Nunnenkamp}, \citenamefont {Ramsay},\ and\ \citenamefont {Ferguson}}]{haigh2016triple}%
	  \BibitemOpen
	  \bibfield  {author} {\bibinfo {author} {\bibfnamefont {J.~A.}\ \bibnamefont {Haigh}}, \bibinfo {author} {\bibfnamefont {A.}~\bibnamefont {Nunnenkamp}}, \bibinfo {author} {\bibfnamefont {A.~J.}\ \bibnamefont {Ramsay}}, \ and\ \bibinfo {author} {\bibfnamefont {A.~J.}\ \bibnamefont {Ferguson}},\ }\bibfield  {title} {\  {\bibinfo {title} {Triple-resonant brillouin light scattering in magneto-optical cavities},\ }}\href {\doibase 10.1103/PhysRevLett.117.133602} {\bibfield  {journal} {\bibinfo  {journal} {Phys. Rev. Lett.}\ }\textbf {\bibinfo {volume} {117}},\ \bibinfo {pages} {133602} (\bibinfo {year} {2016})}\BibitemShut {NoStop}%
	\bibitem [{\citenamefont {Kounalakis}\ \  {et~al.}(2022)\citenamefont {Kounalakis}, \citenamefont {Bauer},\ and\ \citenamefont {Blanter}}]{kounalakis2022analog}%
	  \BibitemOpen
	  \bibfield  {author} {\bibinfo {author} {\bibfnamefont {M.}~\bibnamefont {Kounalakis}}, \bibinfo {author} {\bibfnamefont {G.~E.~W.}\ \bibnamefont {Bauer}}, \ and\ \bibinfo {author} {\bibfnamefont {Y.~M.}\ \bibnamefont {Blanter}},\ }\bibfield  {title} {\  {\bibinfo {title} {Analog quantum control of magnonic cat states on a chip by a superconducting qubit},\ }}\href {\doibase 10.1103/PhysRevLett.129.037205} {\bibfield  {journal} {\bibinfo  {journal} {Phys. Rev. Lett.}\ }\textbf {\bibinfo {volume} {129}},\ \bibinfo {pages} {037205} (\bibinfo {year} {2022})}\BibitemShut {NoStop}%
	\bibitem [{\citenamefont {Wang}\ \  {et~al.}(2018)\citenamefont {Wang}, \citenamefont {Zhang}, \citenamefont {Zhang}, \citenamefont {Li}, \citenamefont {Hu},\ and\ \citenamefont {You}}]{wang2018bistability}%
	  \BibitemOpen
	  \bibfield  {author} {\bibinfo {author} {\bibfnamefont {Y.-P.}\ \bibnamefont {Wang}}, \bibinfo {author} {\bibfnamefont {G.-Q.}\ \bibnamefont {Zhang}}, \bibinfo {author} {\bibfnamefont {D.}~\bibnamefont {Zhang}}, \bibinfo {author} {\bibfnamefont {T.-F.}\ \bibnamefont {Li}}, \bibinfo {author} {\bibfnamefont {C.-M.}\ \bibnamefont {Hu}}, \ and\ \bibinfo {author} {\bibfnamefont {J.~Q.}\ \bibnamefont {You}},\ }\bibfield  {title} {\  {\bibinfo {title} {Bistability of cavity magnon polaritons},\ }}\href {\doibase 10.1103/PhysRevLett.120.057202} {\bibfield  {journal} {\bibinfo  {journal} {Phys. Rev. Lett.}\ }\textbf {\bibinfo {volume} {120}},\ \bibinfo {pages} {057202} (\bibinfo {year} {2018})}\BibitemShut {NoStop}%
	\bibitem [{\citenamefont {Zhang}\ \  {et~al.}(2015)\citenamefont {Zhang}, \citenamefont {Zou}, \citenamefont {Zhu}, \citenamefont {Marquardt}, \citenamefont {Jiang},\ and\ \citenamefont {Tang}}]{zhang2015magnon}%
	  \BibitemOpen
	  \bibfield  {author} {\bibinfo {author} {\bibfnamefont {X.}~\bibnamefont {Zhang}}, \bibinfo {author} {\bibfnamefont {C.-L.}\ \bibnamefont {Zou}}, \bibinfo {author} {\bibfnamefont {N.}~\bibnamefont {Zhu}}, \bibinfo {author} {\bibfnamefont {F.}~\bibnamefont {Marquardt}}, \bibinfo {author} {\bibfnamefont {L.}~\bibnamefont {Jiang}}, \ and\ \bibinfo {author} {\bibfnamefont {H.~X.}\ \bibnamefont {Tang}},\ }\bibfield  {title} {\  {\bibinfo {title} {Magnon dark modes and gradient memory},\ }}\href {https://doi.org/10.1038/ncomms9914} {\bibfield  {journal} {\bibinfo  {journal} {Nat. commun.}\ }\textbf {\bibinfo {volume} {6}},\ \bibinfo {pages} {8914} (\bibinfo {year} {2015})}\BibitemShut {NoStop}%
	\bibitem [{\citenamefont {Xiao}\ \  {et~al.}(2019)\citenamefont {Xiao}, \citenamefont {Yan}, \citenamefont {Zhang}, \citenamefont {Grigoryan}, \citenamefont {Hu}, \citenamefont {Guo},\ and\ \citenamefont {Xia}}]{xiao2019magnon}%
	  \BibitemOpen
	  \bibfield  {author} {\bibinfo {author} {\bibfnamefont {Y.}~\bibnamefont {Xiao}}, \bibinfo {author} {\bibfnamefont {X.~H.}\ \bibnamefont {Yan}}, \bibinfo {author} {\bibfnamefont {Y.}~\bibnamefont {Zhang}}, \bibinfo {author} {\bibfnamefont {V.~L.}\ \bibnamefont {Grigoryan}}, \bibinfo {author} {\bibfnamefont {C.~M.}\ \bibnamefont {Hu}}, \bibinfo {author} {\bibfnamefont {H.}~\bibnamefont {Guo}}, \ and\ \bibinfo {author} {\bibfnamefont {K.}~\bibnamefont {Xia}},\ }\bibfield  {title} {\  {\bibinfo {title} {Magnon dark mode of an antiferromagnetic insulator in a microwave cavity},\ }}\href {\doibase 10.1103/PhysRevB.99.094407} {\bibfield  {journal} {\bibinfo  {journal} {Phys. Rev. B}\ }\textbf {\bibinfo {volume} {99}},\ \bibinfo {pages} {094407} (\bibinfo {year} {2019})}\BibitemShut {NoStop}%
	\bibitem [{\citenamefont {Iguchi}\ \  {et~al.}(2015)\citenamefont {Iguchi}, \citenamefont {Uemura}, \citenamefont {Ueno},\ and\ \citenamefont {Onose}}]{iguchi2015nonreciprocal}%
	  \BibitemOpen
	  \bibfield  {author} {\bibinfo {author} {\bibfnamefont {Y.}~\bibnamefont {Iguchi}}, \bibinfo {author} {\bibfnamefont {S.}~\bibnamefont {Uemura}}, \bibinfo {author} {\bibfnamefont {K.}~\bibnamefont {Ueno}}, \ and\ \bibinfo {author} {\bibfnamefont {Y.}~\bibnamefont {Onose}},\ }\bibfield  {title} {\  {\bibinfo {title} {Nonreciprocal magnon propagation in a noncentrosymmetric ferromagnet ${\text{life}}_{5}{\text{o}}_{8}$},\ }}\href {\doibase 10.1103/PhysRevB.92.184419} {\bibfield  {journal} {\bibinfo  {journal} {Phys. Rev. B}\ }\textbf {\bibinfo {volume} {92}},\ \bibinfo {pages} {184419} (\bibinfo {year} {2015})}\BibitemShut {NoStop}%
	\bibitem [{\citenamefont {Kong}\ \  {et~al.}(2019)\citenamefont {Kong}, \citenamefont {Xiong},\ and\ \citenamefont {Wu}}]{kong2019magnon}%
	  \BibitemOpen
	  \bibfield  {author} {\bibinfo {author} {\bibfnamefont {C.}~\bibnamefont {Kong}}, \bibinfo {author} {\bibfnamefont {H.}~\bibnamefont {Xiong}}, \ and\ \bibinfo {author} {\bibfnamefont {Y.}~\bibnamefont {Wu}},\ }\bibfield  {title} {\  {\bibinfo {title} {Magnon-induced nonreciprocity based on the magnon kerr effect},\ }}\href {\doibase 10.1103/PhysRevApplied.12.034001} {\bibfield  {journal} {\bibinfo  {journal} {Phys. Rev. Appl.}\ }\textbf {\bibinfo {volume} {12}},\ \bibinfo {pages} {034001} (\bibinfo {year} {2019})}\BibitemShut {NoStop}%
	\bibitem [{\citenamefont {Wang}\ \  {et~al.}(2016)\citenamefont {Wang}, \citenamefont {Zhang}, \citenamefont {Zhang}, \citenamefont {Luo}, \citenamefont {Xiong}, \citenamefont {Wang}, \citenamefont {Li}, \citenamefont {Hu},\ and\ \citenamefont {You}}]{wang2016magnon}%
	  \BibitemOpen
	  \bibfield  {author} {\bibinfo {author} {\bibfnamefont {Y.-P.}\ \bibnamefont {Wang}}, \bibinfo {author} {\bibfnamefont {G.-Q.}\ \bibnamefont {Zhang}}, \bibinfo {author} {\bibfnamefont {D.}~\bibnamefont {Zhang}}, \bibinfo {author} {\bibfnamefont {X.-Q.}\ \bibnamefont {Luo}}, \bibinfo {author} {\bibfnamefont {W.}~\bibnamefont {Xiong}}, \bibinfo {author} {\bibfnamefont {S.-P.}\ \bibnamefont {Wang}}, \bibinfo {author} {\bibfnamefont {T.-F.}\ \bibnamefont {Li}}, \bibinfo {author} {\bibfnamefont {C.-M.}\ \bibnamefont {Hu}}, \ and\ \bibinfo {author} {\bibfnamefont {J.~Q.}\ \bibnamefont {You}},\ }\bibfield  {title} {\  {\bibinfo {title} {Magnon kerr effect in a strongly coupled cavity-magnon system},\ }}\href {\doibase 10.1103/PhysRevB.94.224410} {\bibfield  {journal} {\bibinfo  {journal} {Phys. Rev. B}\ }\textbf {\bibinfo {volume} {94}},\ \bibinfo {pages} {224410} (\bibinfo {year} {2016})}\BibitemShut {NoStop}%
	\bibitem [{\citenamefont {Sharma}\ \  {et~al.}(2018)\citenamefont {Sharma}, \citenamefont {Blanter},\ and\ \citenamefont {Bauer}}]{sharma2018optical}%
	  \BibitemOpen
	  \bibfield  {author} {\bibinfo {author} {\bibfnamefont {S.}~\bibnamefont {Sharma}}, \bibinfo {author} {\bibfnamefont {Y.~M.}\ \bibnamefont {Blanter}}, \ and\ \bibinfo {author} {\bibfnamefont {G.~E.~W.}\ \bibnamefont {Bauer}},\ }\bibfield  {title} {\  {\bibinfo {title} {Optical cooling of magnons},\ }}\href {\doibase 10.1103/PhysRevLett.121.087205} {\bibfield  {journal} {\bibinfo  {journal} {Phys. Rev. Lett.}\ }\textbf {\bibinfo {volume} {121}},\ \bibinfo {pages} {087205} (\bibinfo {year} {2018})}\BibitemShut {NoStop}%
	\bibitem [{\citenamefont {Zhang}\ \  {et~al.}(2021)\citenamefont {Zhang}, \citenamefont {Wang}, \citenamefont {Bai}, \citenamefont {Wang}, \citenamefont {Zhang},\ and\ \citenamefont {Wang}}]{zhang2021generation}%
	  \BibitemOpen
	  \bibfield  {author} {\bibinfo {author} {\bibfnamefont {W.}~\bibnamefont {Zhang}}, \bibinfo {author} {\bibfnamefont {D.-Y.}\ \bibnamefont {Wang}}, \bibinfo {author} {\bibfnamefont {C.-H.}\ \bibnamefont {Bai}}, \bibinfo {author} {\bibfnamefont {T.}~\bibnamefont {Wang}}, \bibinfo {author} {\bibfnamefont {S.}~\bibnamefont {Zhang}}, \ and\ \bibinfo {author} {\bibfnamefont {H.-F.}\ \bibnamefont {Wang}},\ }\bibfield  {title} {\  {\bibinfo {title} {Generation and transfer of squeezed states in a cavity magnomechanical system by two-tone microwave fields},\ }}\href {\doibase 10.1364/OE.418531} {\bibfield  {journal} {\bibinfo  {journal} {Opt. Express}\ }\textbf {\bibinfo {volume} {29}},\ \bibinfo {pages} {11773} (\bibinfo {year} {2021})}\BibitemShut {NoStop}%
	\bibitem [{\citenamefont {Li}\ \  {et~al.}(2019{\natexlab{a}})\citenamefont {Li}, \citenamefont {Zhu},\ and\ \citenamefont {Agarwal}}]{li2019squeezed}%
	  \BibitemOpen
	  \bibfield  {author} {\bibinfo {author} {\bibfnamefont {J.}~\bibnamefont {Li}}, \bibinfo {author} {\bibfnamefont {S.-Y.}\ \bibnamefont {Zhu}}, \ and\ \bibinfo {author} {\bibfnamefont {G.~S.}\ \bibnamefont {Agarwal}},\ }\bibfield  {title} {\  {\bibinfo {title} {Squeezed states of magnons and phonons in cavity magnomechanics},\ }}\href {\doibase 10.1103/PhysRevA.99.021801} {\bibfield  {journal} {\bibinfo  {journal} {Phys. Rev. A}\ }\textbf {\bibinfo {volume} {99}},\ \bibinfo {pages} {021801} (\bibinfo {year} {2019}{\natexlab{a}})}\BibitemShut {NoStop}%
	\bibitem [{\citenamefont {Aharonovich}\ \  {et~al.}(2011)\citenamefont {Aharonovich}, \citenamefont {Greentree},\ and\ \citenamefont {Prawer}}]{aharonovich2011diamond}%
	  \BibitemOpen
	  \bibfield  {author} {\bibinfo {author} {\bibfnamefont {I.}~\bibnamefont {Aharonovich}}, \bibinfo {author} {\bibfnamefont {A.~D.}\ \bibnamefont {Greentree}}, \ and\ \bibinfo {author} {\bibfnamefont {S.}~\bibnamefont {Prawer}},\ }\bibfield  {title} {\  {\bibinfo {title} {Diamond photonics},\ }}\href {https://doi.org/10.1038/nphoton.2011.54} {\bibfield  {journal} {\bibinfo  {journal} {Nat. Photon.}\ }\textbf {\bibinfo {volume} {5}},\ \bibinfo {pages} {397} (\bibinfo {year} {2011})}\BibitemShut {NoStop}%
	\bibitem [{\citenamefont {Barry}\ \  {et~al.}(2020)\citenamefont {Barry}, \citenamefont {Schloss}, \citenamefont {Bauch}, \citenamefont {Turner}, \citenamefont {Hart}, \citenamefont {Pham},\ and\ \citenamefont {Walsworth}}]{barry2020sensitivity}%
	  \BibitemOpen
	  \bibfield  {author} {\bibinfo {author} {\bibfnamefont {J.~F.}\ \bibnamefont {Barry}}, \bibinfo {author} {\bibfnamefont {J.~M.}\ \bibnamefont {Schloss}}, \bibinfo {author} {\bibfnamefont {E.}~\bibnamefont {Bauch}}, \bibinfo {author} {\bibfnamefont {M.~J.}\ \bibnamefont {Turner}}, \bibinfo {author} {\bibfnamefont {C.~A.}\ \bibnamefont {Hart}}, \bibinfo {author} {\bibfnamefont {L.~M.}\ \bibnamefont {Pham}}, \ and\ \bibinfo {author} {\bibfnamefont {R.~L.}\ \bibnamefont {Walsworth}},\ }\bibfield  {title} {\  {\bibinfo {title} {Sensitivity optimization for nv-diamond magnetometry},\ }}\href {\doibase 10.1103/RevModPhys.92.015004} {\bibfield  {journal} {\bibinfo  {journal} {Rev. Mod. Phys.}\ }\textbf {\bibinfo {volume} {92}},\ \bibinfo {pages} {015004} (\bibinfo {year} {2020})}\BibitemShut {NoStop}%
	\bibitem [{\citenamefont {Doherty}\ \  {et~al.}(2013)\citenamefont {Doherty}, \citenamefont {Manson}, \citenamefont {Delaney}, \citenamefont {Jelezko}, \citenamefont {Wrachtrup},\ and\ \citenamefont {Hollenberg}}]{doherty2013nitrogen}%
	  \BibitemOpen
	  \bibfield  {author} {\bibinfo {author} {\bibfnamefont {M.~W.}\ \bibnamefont {Doherty}}, \bibinfo {author} {\bibfnamefont {N.~B.}\ \bibnamefont {Manson}}, \bibinfo {author} {\bibfnamefont {P.}~\bibnamefont {Delaney}}, \bibinfo {author} {\bibfnamefont {F.}~\bibnamefont {Jelezko}}, \bibinfo {author} {\bibfnamefont {J.}~\bibnamefont {Wrachtrup}}, \ and\ \bibinfo {author} {\bibfnamefont {L.~C.}\ \bibnamefont {Hollenberg}},\ }\bibfield  {title} {\  {\bibinfo {title} {The nitrogen-vacancy colour centre in diamond},\ }}\href {\doibase https://doi.org/10.1016/j.physrep.2013.02.001} {\bibfield  {journal} {\bibinfo  {journal} {Phys. Rep.}\ }\textbf {\bibinfo {volume} {528}},\ \bibinfo {pages} {1} (\bibinfo {year} {2013})}\BibitemShut {NoStop}%
	\bibitem [{\citenamefont {Bar-Gill}\ \  {et~al.}(2013)\citenamefont {Bar-Gill}, \citenamefont {Pham}, \citenamefont {Jarmola}, \citenamefont {Budker},\ and\ \citenamefont {Walsworth}}]{bar2013solid}%
	  \BibitemOpen
	  \bibfield  {author} {\bibinfo {author} {\bibfnamefont {N.}~\bibnamefont {Bar-Gill}}, \bibinfo {author} {\bibfnamefont {L.~M.}\ \bibnamefont {Pham}}, \bibinfo {author} {\bibfnamefont {A.}~\bibnamefont {Jarmola}}, \bibinfo {author} {\bibfnamefont {D.}~\bibnamefont {Budker}}, \ and\ \bibinfo {author} {\bibfnamefont {R.~L.}\ \bibnamefont {Walsworth}},\ }\bibfield  {title} {\  {\bibinfo {title} {Solid-state electronic spin coherence time approaching one second},\ }}\href {https://doi.org/10.1038/ncomms2771} {\bibfield  {journal} {\bibinfo  {journal} {Nat. Commun.}\ }\textbf {\bibinfo {volume} {4}},\ \bibinfo {pages} {1743} (\bibinfo {year} {2013})}\BibitemShut {NoStop}%
	\bibitem [{\citenamefont {Doherty}\ \  {et~al.}(2014)\citenamefont {Doherty}, \citenamefont {Struzhkin}, \citenamefont {Simpson}, \citenamefont {McGuinness}, \citenamefont {Meng}, \citenamefont {Stacey}, \citenamefont {Karle}, \citenamefont {Hemley}, \citenamefont {Manson}, \citenamefont {Hollenberg},\ and\ \citenamefont {Prawer}}]{doherty2014electronic}%
	  \BibitemOpen
	  \bibfield  {author} {\bibinfo {author} {\bibfnamefont {M.~W.}\ \bibnamefont {Doherty}}, \bibinfo {author} {\bibfnamefont {V.~V.}\ \bibnamefont {Struzhkin}}, \bibinfo {author} {\bibfnamefont {D.~A.}\ \bibnamefont {Simpson}}, \bibinfo {author} {\bibfnamefont {L.~P.}\ \bibnamefont {McGuinness}}, \bibinfo {author} {\bibfnamefont {Y.}~\bibnamefont {Meng}}, \bibinfo {author} {\bibfnamefont {A.}~\bibnamefont {Stacey}}, \bibinfo {author} {\bibfnamefont {T.~J.}\ \bibnamefont {Karle}}, \bibinfo {author} {\bibfnamefont {R.~J.}\ \bibnamefont {Hemley}}, \bibinfo {author} {\bibfnamefont {N.~B.}\ \bibnamefont {Manson}}, \bibinfo {author} {\bibfnamefont {L.~C.~L.}\ \bibnamefont {Hollenberg}}, \ and\ \bibinfo {author} {\bibfnamefont {S.}~\bibnamefont {Prawer}},\ }\bibfield  {title} {\  {\bibinfo {title} {Electronic properties and metrology applications of the diamond ${\mathrm{nv}}^{\ensuremath{-}}$ center under pressure},\ }}\href {\doibase 10.1103/PhysRevLett.112.047601} {\bibfield  {journal} {\bibinfo  {journal} {Phys. Rev. Lett.}\ }\textbf {\bibinfo {volume} {112}},\ \bibinfo {pages} {047601} (\bibinfo {year} {2014})}\BibitemShut {NoStop}%
	\bibitem [{\citenamefont {Abobeih}\ \  {et~al.}(2018)\citenamefont {Abobeih}, \citenamefont {Cramer}, \citenamefont {Bakker}, \citenamefont {Kalb}, \citenamefont {Markham}, \citenamefont {Twitchen},\ and\ \citenamefont {Taminiau}}]{abobeih2018one}%
	  \BibitemOpen
	  \bibfield  {author} {\bibinfo {author} {\bibfnamefont {M.~H.}\ \bibnamefont {Abobeih}}, \bibinfo {author} {\bibfnamefont {J.}~\bibnamefont {Cramer}}, \bibinfo {author} {\bibfnamefont {M.~A.}\ \bibnamefont {Bakker}}, \bibinfo {author} {\bibfnamefont {N.}~\bibnamefont {Kalb}}, \bibinfo {author} {\bibfnamefont {M.}~\bibnamefont {Markham}}, \bibinfo {author} {\bibfnamefont {D.~J.}\ \bibnamefont {Twitchen}}, \ and\ \bibinfo {author} {\bibfnamefont {T.~H.}\ \bibnamefont {Taminiau}},\ }\bibfield  {title} {\  {\bibinfo {title} {One-second coherence for a single electron spin coupled to a multi-qubit nuclear-spin environment},\ }}\href {https://doi.org/10.1038/s41467-018-04916-z} {\bibfield  {journal} {\bibinfo  {journal} {Nat. Commun.}\ }\textbf {\bibinfo {volume} {9}},\ \bibinfo {pages} {2552} (\bibinfo {year} {2018})}\BibitemShut {NoStop}%
	\bibitem [{\citenamefont {Fuchs}\ \  {et~al.}(2011)\citenamefont {Fuchs}, \citenamefont {Burkard}, \citenamefont {Klimov},\ and\ \citenamefont {Awschalom}}]{fuchs2011quantum}%
	  \BibitemOpen
	  \bibfield  {author} {\bibinfo {author} {\bibfnamefont {G.}~\bibnamefont {Fuchs}}, \bibinfo {author} {\bibfnamefont {G.}~\bibnamefont {Burkard}}, \bibinfo {author} {\bibfnamefont {P.}~\bibnamefont {Klimov}}, \ and\ \bibinfo {author} {\bibfnamefont {D.}~\bibnamefont {Awschalom}},\ }\bibfield  {title} {\  {\bibinfo {title} {A quantum memory intrinsic to single nitrogen--vacancy centres in diamond},\ }}\href {https://doi.org/10.1038/nphys2026} {\bibfield  {journal} {\bibinfo  {journal} {Nat. Phys.}\ }\textbf {\bibinfo {volume} {7}},\ \bibinfo {pages} {789} (\bibinfo {year} {2011})}\BibitemShut {NoStop}%
	\bibitem [{\citenamefont {Kolkowitz}\ \  {et~al.}(2012)\citenamefont {Kolkowitz}, \citenamefont {Jayich}, \citenamefont {Unterreithmeier}, \citenamefont {Bennett}, \citenamefont {Rabl}, \citenamefont {Harris},\ and\ \citenamefont {Lukin}}]{kolkowitz2012coherent}%
	  \BibitemOpen
	  \bibfield  {author} {\bibinfo {author} {\bibfnamefont {S.}~\bibnamefont {Kolkowitz}}, \bibinfo {author} {\bibfnamefont {A.~C.~B.}\ \bibnamefont {Jayich}}, \bibinfo {author} {\bibfnamefont {Q.~P.}\ \bibnamefont {Unterreithmeier}}, \bibinfo {author} {\bibfnamefont {S.~D.}\ \bibnamefont {Bennett}}, \bibinfo {author} {\bibfnamefont {P.}~\bibnamefont {Rabl}}, \bibinfo {author} {\bibfnamefont {J.~G.~E.}\ \bibnamefont {Harris}}, \ and\ \bibinfo {author} {\bibfnamefont {M.~D.}\ \bibnamefont {Lukin}},\ }\bibfield  {title} {\  {\bibinfo {title} {Coherent sensing of a mechanical resonator with a single-spin qubit},\ }}\href {\doibase 10.1126/science.1216821} {\bibfield  {journal} {\bibinfo  {journal} {Science}\ }\textbf {\bibinfo {volume} {335}},\ \bibinfo {pages} {1603} (\bibinfo {year} {2012})}\BibitemShut {NoStop}%
	\bibitem [{\citenamefont {Dolde}\ \  {et~al.}(2011)\citenamefont {Dolde}, \citenamefont {Fedder}, \citenamefont {Doherty}, \citenamefont {N{\"o}bauer}, \citenamefont {Rempp}, \citenamefont {Balasubramanian}, \citenamefont {Wolf}, \citenamefont {Reinhard}, \citenamefont {Hollenberg}, \citenamefont {Jelezko} \  {et~al.}}]{dolde2011electric}%
	  \BibitemOpen
	  \bibfield  {author} {\bibinfo {author} {\bibfnamefont {F.}~\bibnamefont {Dolde}}, \bibinfo {author} {\bibfnamefont {H.}~\bibnamefont {Fedder}}, \bibinfo {author} {\bibfnamefont {M.~W.}\ \bibnamefont {Doherty}}, \bibinfo {author} {\bibfnamefont {T.}~\bibnamefont {N{\"o}bauer}}, \bibinfo {author} {\bibfnamefont {F.}~\bibnamefont {Rempp}}, \bibinfo {author} {\bibfnamefont {G.}~\bibnamefont {Balasubramanian}}, \bibinfo {author} {\bibfnamefont {T.}~\bibnamefont {Wolf}}, \bibinfo {author} {\bibfnamefont {F.}~\bibnamefont {Reinhard}}, \bibinfo {author} {\bibfnamefont {L.~C.}\ \bibnamefont {Hollenberg}}, \bibinfo {author} {\bibfnamefont {F.}~\bibnamefont {Jelezko}},  \  {et~al.},\ }\bibfield  {title} {\  {\bibinfo {title} {Electric-field sensing using single diamond spins},\ }}\href {https://doi.org/10.1038/nphys1969} {\bibfield  {journal} {\bibinfo  {journal} {Nat. Phys.}\ }\textbf {\bibinfo {volume} {7}},\ \bibinfo {pages} {459} (\bibinfo {year} {2011})}\BibitemShut {NoStop}%
	\bibitem [{\citenamefont {Birnbaum}\ \  {et~al.}(2005)\citenamefont {Birnbaum}, \citenamefont {Boca}, \citenamefont {Miller}, \citenamefont {Boozer}, \citenamefont {Northup},\ and\ \citenamefont {Kimble}}]{birnbaum2005photon}%
	  \BibitemOpen
	  \bibfield  {author} {\bibinfo {author} {\bibfnamefont {K.~M.}\ \bibnamefont {Birnbaum}}, \bibinfo {author} {\bibfnamefont {A.}~\bibnamefont {Boca}}, \bibinfo {author} {\bibfnamefont {R.}~\bibnamefont {Miller}}, \bibinfo {author} {\bibfnamefont {A.~D.}\ \bibnamefont {Boozer}}, \bibinfo {author} {\bibfnamefont {T.~E.}\ \bibnamefont {Northup}}, \ and\ \bibinfo {author} {\bibfnamefont {H.~J.}\ \bibnamefont {Kimble}},\ }\bibfield  {title} {\  {\bibinfo {title} {Photon blockade in an optical cavity with one trapped atom},\ }}\href {https://doi.org/10.1038/nature03804} {\bibfield  {journal} {\bibinfo  {journal} {Nature}\ }\textbf {\bibinfo {volume} {436}},\ \bibinfo {pages} {87} (\bibinfo {year} {2005})}\BibitemShut {NoStop}%
	\bibitem [{\citenamefont {Rabl}(2011)}]{rabl2011photon}%
	  \BibitemOpen
	  \bibfield  {author} {\bibinfo {author} {\bibfnamefont {P.}~\bibnamefont {Rabl}},\ }\bibfield  {title} {\  {\bibinfo {title} {Photon blockade effect in optomechanical systems},\ }}\href {\doibase 10.1103/PhysRevLett.107.063601} {\bibfield  {journal} {\bibinfo  {journal} {Phys. Rev. Lett.}\ }\textbf {\bibinfo {volume} {107}},\ \bibinfo {pages} {063601} (\bibinfo {year} {2011})}\BibitemShut {NoStop}%
	\bibitem [{\citenamefont {Ridolfo}\ \  {et~al.}(2012)\citenamefont {Ridolfo}, \citenamefont {Leib}, \citenamefont {Savasta},\ and\ \citenamefont {Hartmann}}]{ridolfo2012photon}%
	  \BibitemOpen
	  \bibfield  {author} {\bibinfo {author} {\bibfnamefont {A.}~\bibnamefont {Ridolfo}}, \bibinfo {author} {\bibfnamefont {M.}~\bibnamefont {Leib}}, \bibinfo {author} {\bibfnamefont {S.}~\bibnamefont {Savasta}}, \ and\ \bibinfo {author} {\bibfnamefont {M.~J.}\ \bibnamefont {Hartmann}},\ }\bibfield  {title} {\  {\bibinfo {title} {Photon blockade in the ultrastrong coupling regime},\ }}\href {\doibase 10.1103/PhysRevLett.109.193602} {\bibfield  {journal} {\bibinfo  {journal} {Phys. Rev. Lett.}\ }\textbf {\bibinfo {volume} {109}},\ \bibinfo {pages} {193602} (\bibinfo {year} {2012})}\BibitemShut {NoStop}%
	\bibitem [{\citenamefont {Liu}\ \  {et~al.}(2010)\citenamefont {Liu}, \citenamefont {Miranowicz}, \citenamefont {Gao}, \citenamefont {Bajer}, \citenamefont {Sun},\ and\ \citenamefont {Nori}}]{liu2010qubit}%
	  \BibitemOpen
	  \bibfield  {author} {\bibinfo {author} {\bibfnamefont {Y.-x.}\ \bibnamefont {Liu}}, \bibinfo {author} {\bibfnamefont {A.}~\bibnamefont {Miranowicz}}, \bibinfo {author} {\bibfnamefont {Y.~B.}\ \bibnamefont {Gao}}, \bibinfo {author} {\bibfnamefont {J.~c.~v.}\ \bibnamefont {Bajer}}, \bibinfo {author} {\bibfnamefont {C.~P.}\ \bibnamefont {Sun}}, \ and\ \bibinfo {author} {\bibfnamefont {F.}~\bibnamefont {Nori}},\ }\bibfield  {title} {\  {\bibinfo {title} {Qubit-induced phonon blockade as a signature of quantum behavior in nanomechanical resonators},\ }}\href {\doibase 10.1103/PhysRevA.82.032101} {\bibfield  {journal} {\bibinfo  {journal} {Phys. Rev. A}\ }\textbf {\bibinfo {volume} {82}},\ \bibinfo {pages} {032101} (\bibinfo {year} {2010})}\BibitemShut {NoStop}%
	\bibitem [{\citenamefont {Xie}\ \  {et~al.}(2017)\citenamefont {Xie}, \citenamefont {Liao}, \citenamefont {Shang}, \citenamefont {Ye},\ and\ \citenamefont {Lin}}]{xie2017phonon}%
	  \BibitemOpen
	  \bibfield  {author} {\bibinfo {author} {\bibfnamefont {H.}~\bibnamefont {Xie}}, \bibinfo {author} {\bibfnamefont {C.-G.}\ \bibnamefont {Liao}}, \bibinfo {author} {\bibfnamefont {X.}~\bibnamefont {Shang}}, \bibinfo {author} {\bibfnamefont {M.-Y.}\ \bibnamefont {Ye}}, \ and\ \bibinfo {author} {\bibfnamefont {X.-M.}\ \bibnamefont {Lin}},\ }\bibfield  {title} {\  {\bibinfo {title} {Phonon blockade in a quadratically coupled optomechanical system},\ }}\href {\doibase 10.1103/PhysRevA.96.013861} {\bibfield  {journal} {\bibinfo  {journal} {Phys. Rev. A}\ }\textbf {\bibinfo {volume} {96}},\ \bibinfo {pages} {013861} (\bibinfo {year} {2017})}\BibitemShut {NoStop}%
	\bibitem [{\citenamefont {Yao}\ \  {et~al.}(2022)\citenamefont {Yao}, \citenamefont {Ali}, \citenamefont {Li},\ and\ \citenamefont {Li}}]{yao2022nonreciprocal}%
	  \BibitemOpen
	  \bibfield  {author} {\bibinfo {author} {\bibfnamefont {X.-Y.}\ \bibnamefont {Yao}}, \bibinfo {author} {\bibfnamefont {H.}~\bibnamefont {Ali}}, \bibinfo {author} {\bibfnamefont {F.-L.}\ \bibnamefont {Li}}, \ and\ \bibinfo {author} {\bibfnamefont {P.-B.}\ \bibnamefont {Li}},\ }\bibfield  {title} {\  {\bibinfo {title} {Nonreciprocal phonon blockade in a spinning acoustic ring cavity coupled to a two-level system},\ }}\href {\doibase 10.1103/PhysRevApplied.17.054004} {\bibfield  {journal} {\bibinfo  {journal} {Phys. Rev. Appl.}\ }\textbf {\bibinfo {volume} {17}},\ \bibinfo {pages} {054004} (\bibinfo {year} {2022})}\BibitemShut {NoStop}%
	\bibitem [{\citenamefont {Liu}\ \  {et~al.}(2019)\citenamefont {Liu}, \citenamefont {Xiong},\ and\ \citenamefont {Wu}}]{liu2019magnon}%
	  \BibitemOpen
	  \bibfield  {author} {\bibinfo {author} {\bibfnamefont {Z.-X.}\ \bibnamefont {Liu}}, \bibinfo {author} {\bibfnamefont {H.}~\bibnamefont {Xiong}}, \ and\ \bibinfo {author} {\bibfnamefont {Y.}~\bibnamefont {Wu}},\ }\bibfield  {title} {\  {\bibinfo {title} {Magnon blockade in a hybrid ferromagnet-superconductor quantum system},\ }}\href {\doibase 10.1103/PhysRevB.100.134421} {\bibfield  {journal} {\bibinfo  {journal} {Phys. Rev. B}\ }\textbf {\bibinfo {volume} {100}},\ \bibinfo {pages} {134421} (\bibinfo {year} {2019})}\BibitemShut {NoStop}%
	\bibitem [{\citenamefont {Xie}\ \  {et~al.}(2020)\citenamefont {Xie}, \citenamefont {Ma},\ and\ \citenamefont {Li}}]{xie2020quantum}%
	  \BibitemOpen
	  \bibfield  {author} {\bibinfo {author} {\bibfnamefont {J.-k.}\ \bibnamefont {Xie}}, \bibinfo {author} {\bibfnamefont {S.-l.}\ \bibnamefont {Ma}}, \ and\ \bibinfo {author} {\bibfnamefont {F.-l.}\ \bibnamefont {Li}},\ }\bibfield  {title} {\  {\bibinfo {title} {Quantum-interference-enhanced magnon blockade in an yttrium-iron-garnet sphere coupled to superconducting circuits},\ }}\href {\doibase 10.1103/PhysRevA.101.042331} {\bibfield  {journal} {\bibinfo  {journal} {Phys. Rev. A}\ }\textbf {\bibinfo {volume} {101}},\ \bibinfo {pages} {042331} (\bibinfo {year} {2020})}\BibitemShut {NoStop}%
	\bibitem [{\citenamefont {Zhang}\ \  {et~al.}(2024)\citenamefont {Zhang}, \citenamefont {Liu}, \citenamefont {Zhang},\ and\ \citenamefont {Wang}}]{zhang2024magnon}%
	  \BibitemOpen
	  \bibfield  {author} {\bibinfo {author} {\bibfnamefont {W.}~\bibnamefont {Zhang}}, \bibinfo {author} {\bibfnamefont {S.}~\bibnamefont {Liu}}, \bibinfo {author} {\bibfnamefont {S.}~\bibnamefont {Zhang}}, \ and\ \bibinfo {author} {\bibfnamefont {H.-F.}\ \bibnamefont {Wang}},\ }\bibfield  {title} {\  {\bibinfo {title} {Magnon blockade induced by parametric amplification},\ }}\href {\doibase 10.1103/PhysRevA.109.043712} {\bibfield  {journal} {\bibinfo  {journal} {Phys. Rev. A}\ }\textbf {\bibinfo {volume} {109}},\ \bibinfo {pages} {043712} (\bibinfo {year} {2024})}\BibitemShut {NoStop}%
	\bibitem [{\citenamefont {Wang}\ \  {et~al.}(2024)\citenamefont {Wang}, \citenamefont {Huang},\ and\ \citenamefont {Xiong}}]{wang2024magnon}%
	  \BibitemOpen
	  \bibfield  {author} {\bibinfo {author} {\bibfnamefont {X.}~\bibnamefont {Wang}}, \bibinfo {author} {\bibfnamefont {K.-W.}\ \bibnamefont {Huang}}, \ and\ \bibinfo {author} {\bibfnamefont {H.}~\bibnamefont {Xiong}},\ }\bibfield  {title} {\  {\bibinfo {title} {Magnon blockade in a qed system with a giant spin ensemble and a giant atom coupled to a waveguide},\ }}\href {\doibase 10.1103/PhysRevA.110.033702} {\bibfield  {journal} {\bibinfo  {journal} {Phys. Rev. A}\ }\textbf {\bibinfo {volume} {110}},\ \bibinfo {pages} {033702} (\bibinfo {year} {2024})}\BibitemShut {NoStop}%
	\bibitem [{\citenamefont {Yuan}\ \  {et~al.}(2023)\citenamefont {Yuan}, \citenamefont {Chen}, \citenamefont {Han}, \citenamefont {Wu}, \citenamefont {Li}, \citenamefont {Xia}, \citenamefont {Jiang},\ and\ \citenamefont {Song}}]{yuan2023periodic}%
	  \BibitemOpen
	  \bibfield  {author} {\bibinfo {author} {\bibfnamefont {Z.-H.}\ \bibnamefont {Yuan}}, \bibinfo {author} {\bibfnamefont {Y.-J.}\ \bibnamefont {Chen}}, \bibinfo {author} {\bibfnamefont {J.-X.}\ \bibnamefont {Han}}, \bibinfo {author} {\bibfnamefont {J.-L.}\ \bibnamefont {Wu}}, \bibinfo {author} {\bibfnamefont {W.-Q.}\ \bibnamefont {Li}}, \bibinfo {author} {\bibfnamefont {Y.}~\bibnamefont {Xia}}, \bibinfo {author} {\bibfnamefont {Y.-Y.}\ \bibnamefont {Jiang}}, \ and\ \bibinfo {author} {\bibfnamefont {J.}~\bibnamefont {Song}},\ }\bibfield  {title} {\  {\bibinfo {title} {Periodic photon-magnon blockade in an optomagnonic system with chiral exceptional points},\ }}\href {\doibase 10.1103/PhysRevB.108.134409} {\bibfield  {journal} {\bibinfo  {journal} {Phys. Rev. B}\ }\textbf {\bibinfo {volume} {108}},\ \bibinfo {pages} {134409} (\bibinfo {year} {2023})}\BibitemShut {NoStop}%
	\bibitem [{\citenamefont {Jin}\ and\ \citenamefont {Jing}(2023)}]{jin2023magnon}%
	  \BibitemOpen
	  \bibfield  {author} {\bibinfo {author} {\bibfnamefont {Z.-y.}\ \bibnamefont {Jin}}\ and\ \bibinfo {author} {\bibfnamefont {J.}~\bibnamefont {Jing}},\ }\bibfield  {title} {\  {\bibinfo {title} {Magnon blockade in magnon-qubit systems},\ }}\href {\doibase 10.1103/PhysRevA.108.053702} {\bibfield  {journal} {\bibinfo  {journal} {Phys. Rev. A}\ }\textbf {\bibinfo {volume} {108}},\ \bibinfo {pages} {053702} (\bibinfo {year} {2023})}\BibitemShut {NoStop}%
	\bibitem [{\citenamefont {Zhao}\ \  {et~al.}(2020)\citenamefont {Zhao}, \citenamefont {Li}, \citenamefont {Chao}, \citenamefont {Peng}, \citenamefont {Li},\ and\ \citenamefont {Zhou}}]{zhao2020simultaneous}%
	  \BibitemOpen
	  \bibfield  {author} {\bibinfo {author} {\bibfnamefont {C.}~\bibnamefont {Zhao}}, \bibinfo {author} {\bibfnamefont {X.}~\bibnamefont {Li}}, \bibinfo {author} {\bibfnamefont {S.}~\bibnamefont {Chao}}, \bibinfo {author} {\bibfnamefont {R.}~\bibnamefont {Peng}}, \bibinfo {author} {\bibfnamefont {C.}~\bibnamefont {Li}}, \ and\ \bibinfo {author} {\bibfnamefont {L.}~\bibnamefont {Zhou}},\ }\bibfield  {title} {\  {\bibinfo {title} {Simultaneous blockade of a photon, phonon, and magnon induced by a two-level atom},\ }}\href {\doibase 10.1103/PhysRevA.101.063838} {\bibfield  {journal} {\bibinfo  {journal} {Phys. Rev. A}\ }\textbf {\bibinfo {volume} {101}},\ \bibinfo {pages} {063838} (\bibinfo {year} {2020})}\BibitemShut {NoStop}%
	\bibitem [{\citenamefont {Wang}\ \  {et~al.}(2022)\citenamefont {Wang}, \citenamefont {Gou}, \citenamefont {Xu},\ and\ \citenamefont {Gong}}]{wang2022hybrid}%
	  \BibitemOpen
	  \bibfield  {author} {\bibinfo {author} {\bibfnamefont {F.}~\bibnamefont {Wang}}, \bibinfo {author} {\bibfnamefont {C.}~\bibnamefont {Gou}}, \bibinfo {author} {\bibfnamefont {J.}~\bibnamefont {Xu}}, \ and\ \bibinfo {author} {\bibfnamefont {C.}~\bibnamefont {Gong}},\ }\bibfield  {title} {\  {\bibinfo {title} {Hybrid magnon-atom entanglement and magnon blockade via quantum interference},\ }}\href {\doibase 10.1103/PhysRevA.106.013705} {\bibfield  {journal} {\bibinfo  {journal} {Phys. Rev. A}\ }\textbf {\bibinfo {volume} {106}},\ \bibinfo {pages} {013705} (\bibinfo {year} {2022})}\BibitemShut {NoStop}%
	\bibitem [{\citenamefont {Hou}\ \  {et~al.}(2024)\citenamefont {Hou}, \citenamefont {Zhang}, \citenamefont {Han}, \citenamefont {Wang},\ and\ \citenamefont {Zhang}}]{hou2024magnon}%
	  \BibitemOpen
	  \bibfield  {author} {\bibinfo {author} {\bibfnamefont {R.}~\bibnamefont {Hou}}, \bibinfo {author} {\bibfnamefont {W.}~\bibnamefont {Zhang}}, \bibinfo {author} {\bibfnamefont {X.}~\bibnamefont {Han}}, \bibinfo {author} {\bibfnamefont {H.-F.}\ \bibnamefont {Wang}}, \ and\ \bibinfo {author} {\bibfnamefont {S.}~\bibnamefont {Zhang}},\ }\bibfield  {title} {\  {\bibinfo {title} {Magnon blockade based on the kerr nonlinearity in cavity electromagnonics},\ }}\href {\doibase 10.1103/PhysRevA.109.033721} {\bibfield  {journal} {\bibinfo  {journal} {Phys. Rev. A}\ }\textbf {\bibinfo {volume} {109}},\ \bibinfo {pages} {033721} (\bibinfo {year} {2024})}\BibitemShut {NoStop}%
	\bibitem [{\citenamefont {Hei}\ \  {et~al.}(2021)\citenamefont {Hei}, \citenamefont {Dong}, \citenamefont {Chen}, \citenamefont {Shen}, \citenamefont {Qiao},\ and\ \citenamefont {Li}}]{hei2021enhancing}%
	  \BibitemOpen
	  \bibfield  {author} {\bibinfo {author} {\bibfnamefont {X.-L.}\ \bibnamefont {Hei}}, \bibinfo {author} {\bibfnamefont {X.-L.}\ \bibnamefont {Dong}}, \bibinfo {author} {\bibfnamefont {J.-Q.}\ \bibnamefont {Chen}}, \bibinfo {author} {\bibfnamefont {C.-P.}\ \bibnamefont {Shen}}, \bibinfo {author} {\bibfnamefont {Y.-F.}\ \bibnamefont {Qiao}}, \ and\ \bibinfo {author} {\bibfnamefont {P.-B.}\ \bibnamefont {Li}},\ }\bibfield  {title} {\  {\bibinfo {title} {Enhancing spin-photon coupling with a micromagnet},\ }}\href {\doibase 10.1103/PhysRevA.103.043706} {\bibfield  {journal} {\bibinfo  {journal} {Phys. Rev. A}\ }\textbf {\bibinfo {volume} {103}},\ \bibinfo {pages} {043706} (\bibinfo {year} {2021})}\BibitemShut {NoStop}%
	\bibitem [{\citenamefont {Gonzalez-Ballestero}\ \  {et~al.}(2020)\citenamefont {Gonzalez-Ballestero}, \citenamefont {H\"ummer}, \citenamefont {Gieseler},\ and\ \citenamefont {Romero-Isart}}]{gonzalez2020theory}%
	  \BibitemOpen
	  \bibfield  {author} {\bibinfo {author} {\bibfnamefont {C.}~\bibnamefont {Gonzalez-Ballestero}}, \bibinfo {author} {\bibfnamefont {D.}~\bibnamefont {H\"ummer}}, \bibinfo {author} {\bibfnamefont {J.}~\bibnamefont {Gieseler}}, \ and\ \bibinfo {author} {\bibfnamefont {O.}~\bibnamefont {Romero-Isart}},\ }\bibfield  {title} {\  {\bibinfo {title} {Theory of quantum acoustomagnonics and acoustomechanics with a micromagnet},\ }}\href {\doibase 10.1103/PhysRevB.101.125404} {\bibfield  {journal} {\bibinfo  {journal} {Phys. Rev. B}\ }\textbf {\bibinfo {volume} {101}},\ \bibinfo {pages} {125404} (\bibinfo {year} {2020})}\BibitemShut {NoStop}%
	\bibitem [{\citenamefont {Kittel}(1948)}]{kittel1948theory}%
	  \BibitemOpen
	  \bibfield  {author} {\bibinfo {author} {\bibfnamefont {C.}~\bibnamefont {Kittel}},\ }\bibfield  {title} {\  {\bibinfo {title} {On the theory of ferromagnetic resonance absorption},\ }}\href {\doibase 10.1103/PhysRev.73.155} {\bibfield  {journal} {\bibinfo  {journal} {Phys. Rev.}\ }\textbf {\bibinfo {volume} {73}},\ \bibinfo {pages} {155} (\bibinfo {year} {1948})}\BibitemShut {NoStop}%
	\bibitem [{\citenamefont {Tabuchi}\ \  {et~al.}(2016)\citenamefont {Tabuchi}, \citenamefont {Ishino}, \citenamefont {Noguchi}, \citenamefont {Ishikawa}, \citenamefont {Yamazaki}, \citenamefont {Usami},\ and\ \citenamefont {Nakamura}}]{tabuchi2016quantum}%
	  \BibitemOpen
	  \bibfield  {author} {\bibinfo {author} {\bibfnamefont {Y.}~\bibnamefont {Tabuchi}}, \bibinfo {author} {\bibfnamefont {S.}~\bibnamefont {Ishino}}, \bibinfo {author} {\bibfnamefont {A.}~\bibnamefont {Noguchi}}, \bibinfo {author} {\bibfnamefont {T.}~\bibnamefont {Ishikawa}}, \bibinfo {author} {\bibfnamefont {R.}~\bibnamefont {Yamazaki}}, \bibinfo {author} {\bibfnamefont {K.}~\bibnamefont {Usami}}, \ and\ \bibinfo {author} {\bibfnamefont {Y.}~\bibnamefont {Nakamura}},\ }\bibfield  {title} {\  {\bibinfo {title} {Quantum magnonics: The magnon meets the superconducting qubit},\ }}\href {https://doi.org/10.1016/j.crhy.2016.07.009} {\bibfield  {journal} {\bibinfo  {journal} {C. R. Phys.}\ }\textbf {\bibinfo {volume} {17}},\ \bibinfo {pages} {729} (\bibinfo {year} {2016})}\BibitemShut {NoStop}%
	\bibitem [{\citenamefont {Li}\ \  {et~al.}(2019{\natexlab{b}})\citenamefont {Li}, \citenamefont {Li}, \citenamefont {Zhou}, \citenamefont {Liu}, \citenamefont {Li},\ and\ \citenamefont {Li}}]{li2019interfacing}%
	  \BibitemOpen
	  \bibfield  {author} {\bibinfo {author} {\bibfnamefont {B.}~\bibnamefont {Li}}, \bibinfo {author} {\bibfnamefont {P.-B.}\ \bibnamefont {Li}}, \bibinfo {author} {\bibfnamefont {Y.}~\bibnamefont {Zhou}}, \bibinfo {author} {\bibfnamefont {J.}~\bibnamefont {Liu}}, \bibinfo {author} {\bibfnamefont {H.-R.}\ \bibnamefont {Li}}, \ and\ \bibinfo {author} {\bibfnamefont {F.-L.}\ \bibnamefont {Li}},\ }\bibfield  {title} {\  {\bibinfo {title} {Interfacing a topological qubit with a spin qubit in a hybrid quantum system},\ }}\href {\doibase 10.1103/PhysRevApplied.11.044026} {\bibfield  {journal} {\bibinfo  {journal} {Phys. Rev. Appl.}\ }\textbf {\bibinfo {volume} {11}},\ \bibinfo {pages} {044026} (\bibinfo {year} {2019}{\natexlab{b}})}\BibitemShut {NoStop}%
	\bibitem [{\citenamefont {Johansson}\ \  {et~al.}(2012)\citenamefont {Johansson}, \citenamefont {Nation},\ and\ \citenamefont {Nori}}]{johansson2012qutip}%
	  \BibitemOpen
	  \bibfield  {author} {\bibinfo {author} {\bibfnamefont {J.~R.}\ \bibnamefont {Johansson}}, \bibinfo {author} {\bibfnamefont {P.~D.}\ \bibnamefont {Nation}}, \ and\ \bibinfo {author} {\bibfnamefont {F.}~\bibnamefont {Nori}},\ }\bibfield  {title} {\  {\bibinfo {title} {Qutip: An open-source python framework for the dynamics of open quantum systems},\ }}\href {https://doi.org/10.1016/j.cpc.2012.02.021} {\bibfield  {journal} {\bibinfo  {journal} {Comput. phys. commun.}\ }\textbf {\bibinfo {volume} {183}},\ \bibinfo {pages} {1760} (\bibinfo {year} {2012})}\BibitemShut {NoStop}%
	\bibitem [{\citenamefont {Xu}\ \  {et~al.}(2013)\citenamefont {Xu}, \citenamefont {Li},\ and\ \citenamefont {Liu}}]{xu2013photon}%
	  \BibitemOpen
	  \bibfield  {author} {\bibinfo {author} {\bibfnamefont {X.-W.}\ \bibnamefont {Xu}}, \bibinfo {author} {\bibfnamefont {Y.-J.}\ \bibnamefont {Li}}, \ and\ \bibinfo {author} {\bibfnamefont {Y.-x.}\ \bibnamefont {Liu}},\ }\bibfield  {title} {\  {\bibinfo {title} {Photon-induced tunneling in optomechanical systems},\ }}\href {\doibase 10.1103/PhysRevA.87.025803} {\bibfield  {journal} {\bibinfo  {journal} {Phys. Rev. A}\ }\textbf {\bibinfo {volume} {87}},\ \bibinfo {pages} {025803} (\bibinfo {year} {2013})}\BibitemShut {NoStop}%
	\bibitem [{\citenamefont {Kowalewska-Kud\l{}aszyk}\ \  {et~al.}(2019)\citenamefont {Kowalewska-Kud\l{}aszyk}, \citenamefont {Abo}, \citenamefont {Chimczak}, \citenamefont {Pe\ifmmode~\check{r}\else \v{r}\fi{}ina}, \citenamefont {Nori},\ and\ \citenamefont {Miranowicz}}]{kowalewska2019two}%
	  \BibitemOpen
	  \bibfield  {author} {\bibinfo {author} {\bibfnamefont {A.}~\bibnamefont {Kowalewska-Kud\l{}aszyk}}, \bibinfo {author} {\bibfnamefont {S.~I.}\ \bibnamefont {Abo}}, \bibinfo {author} {\bibfnamefont {G.}~\bibnamefont {Chimczak}}, \bibinfo {author} {\bibfnamefont {J.}~\bibnamefont {Pe\ifmmode~\check{r}\else \v{r}\fi{}ina}}, \bibinfo {author} {\bibfnamefont {F.}~\bibnamefont {Nori}}, \ and\ \bibinfo {author} {\bibfnamefont {A.}~\bibnamefont {Miranowicz}},\ }\bibfield  {title} {\  {\bibinfo {title} {Two-photon blockade and photon-induced tunneling generated by squeezing},\ }}\href {\doibase 10.1103/PhysRevA.100.053857} {\bibfield  {journal} {\bibinfo  {journal} {Phys. Rev. A}\ }\textbf {\bibinfo {volume} {100}},\ \bibinfo {pages} {053857} (\bibinfo {year} {2019})}\BibitemShut {NoStop}%
	\bibitem [{\citenamefont {Zhai}\ \  {et~al.}(2019)\citenamefont {Zhai}, \citenamefont {Huang}, \citenamefont {Jing},\ and\ \citenamefont {Kuang}}]{zhai2019mechanical}%
	  \BibitemOpen
	  \bibfield  {author} {\bibinfo {author} {\bibfnamefont {C.}~\bibnamefont {Zhai}}, \bibinfo {author} {\bibfnamefont {R.}~\bibnamefont {Huang}}, \bibinfo {author} {\bibfnamefont {H.}~\bibnamefont {Jing}}, \ and\ \bibinfo {author} {\bibfnamefont {L.-M.}\ \bibnamefont {Kuang}},\ }\bibfield  {title} {\  {\bibinfo {title} {Mechanical switch of photon blockade and photon-induced tunneling},\ }}\href {https://opg.optica.org/oe/abstract.cfm?URI=oe-27-20-27649} {\bibfield  {journal} {\bibinfo  {journal} {Opt. Express}\ }\textbf {\bibinfo {volume} {27}},\ \bibinfo {pages} {27649} (\bibinfo {year} {2019})}\BibitemShut {NoStop}%
	\bibitem [{\citenamefont {Plenio}\ and\ \citenamefont {Knight}(1998)}]{plenio1998quantum}%
	  \BibitemOpen
	  \bibfield  {author} {\bibinfo {author} {\bibfnamefont {M.~B.}\ \bibnamefont {Plenio}}\ and\ \bibinfo {author} {\bibfnamefont {P.~L.}\ \bibnamefont {Knight}},\ }\bibfield  {title} {\  {\bibinfo {title} {The quantum-jump approach to dissipative dynamics in quantum optics},\ }}\href {\doibase 10.1103/RevModPhys.70.101} {\bibfield  {journal} {\bibinfo  {journal} {Rev. Mod. Phys.}\ }\textbf {\bibinfo {volume} {70}},\ \bibinfo {pages} {101} (\bibinfo {year} {1998})}\BibitemShut {NoStop}%
	\bibitem [{\citenamefont {Minganti}\ \  {et~al.}(2019)\citenamefont {Minganti}, \citenamefont {Miranowicz}, \citenamefont {Chhajlany},\ and\ \citenamefont {Nori}}]{minganti2019quantum}%
	  \BibitemOpen
	  \bibfield  {author} {\bibinfo {author} {\bibfnamefont {F.}~\bibnamefont {Minganti}}, \bibinfo {author} {\bibfnamefont {A.}~\bibnamefont {Miranowicz}}, \bibinfo {author} {\bibfnamefont {R.~W.}\ \bibnamefont {Chhajlany}}, \ and\ \bibinfo {author} {\bibfnamefont {F.}~\bibnamefont {Nori}},\ }\bibfield  {title} {\  {\bibinfo {title} {Quantum exceptional points of non-hermitian hamiltonians and liouvillians: The effects of quantum jumps},\ }}\href {\doibase 10.1103/PhysRevA.100.062131} {\bibfield  {journal} {\bibinfo  {journal} {Phys. Rev. A}\ }\textbf {\bibinfo {volume} {100}},\ \bibinfo {pages} {062131} (\bibinfo {year} {2019})}\BibitemShut {NoStop}%
	\end{thebibliography}
\end{document}